%% file: main.tex
\documentclass[aps,pra,twocolumn,superscriptaddress,floatfix,longbibliography, nofootinbib]{revtex4-2}
\usepackage{tabularx}
\usepackage{multirow}
\usepackage[normalem]{ulem}
\pdfoutput=1
\input{allcommands.tex}

\input{figures.tex}

\begin{document}

\title{Approximate Quadratization of High-Order Hamiltonians for \\ Combinatorial Quantum Optimization}

\author{Sabina Dr\u{a}goi}
\affiliation{Institute for Theoretical Physics, ETH Z\"{u}rich, Wolfgang-Pauli-Str. 27, 8093 Z\"{u}rich, Switzerland}
\affiliation{IBM Quantum, IBM Research - Zurich, S\"{a}umerstrasse 4, 8803 R\"{u}schlikon, Switzerland}

\author{Alberto Baiardi}
\affiliation{IBM Quantum, IBM Research - Zurich, S\"{a}umerstrasse 4, 8803 R\"{u}schlikon, Switzerland}

\author{Daniel J. Egger}
\affiliation{IBM Quantum, IBM Research - Zurich, S\"{a}umerstrasse 4, 8803 R\"{u}schlikon, Switzerland}

\begin{abstract}
Combinatorial optimization problems have wide-ranging applications in industry and academia. Quantum computers may help solve them by sampling from appropriately prepared Ansatz quantum circuits. However, current quantum computers are limited by their qubit count, connectivity, and noise. This is particularly restrictive when considering optimization problems beyond the quadratic order. Here, we introduce Ansatze based on an approximate quadratization of high-order Hamiltonians which do not incur a qubit overhead. The price paid is a loss in the quality of the noiseless solution. Crucially, these approximations yield shallower Ansatze which are more robust to noise than the standard QAOA one. We show this through simulations with variable noise strengths. Furthermore, we also propose a noise-aware Ansatz design method for quadratic optimization problems. This method implements only a portion of the target Hamiltonian by limiting the number of layers of SWAP gates in the Ansatz. We find that for both problem types, under noise, our approximate implementation of the full problem structure can significantly enhance the solution quality. Our work opens a path to enhance the solution quality that approximate quantum optimization achieves on noisy hardware.
\end{abstract}

\maketitle

\section{\label{sec:sec1}Introduction}

Quantum computing offers a new information-processing framework based on the laws of quantum physics. For certain tasks, quantum computing achieves better performance than the best known classical methods. Famously, Shor's algorithm delivers an exponential speed-up in factoring over state-of-the-art classical algorithms~\cite{shor1994algorithms}. However, this remains a theoretical feat since Shor's algorithm requires deep circuits which cannot be faithfully implemented at practically relevant scales on current hardware due to noise. This has motivated the development of algorithms requiring shallower circuits such as Variational Quantum Algorithms (VQAs)~\cite{peruzzo2014variational}. VQAs are typically heuristic in nature and are often designed to find the ground state of a Hamiltonian~\cite{tilly2022variational, cao2018potential, blunt2022perspective}. VQAs optimize a parametrized quantum circuit, called the Ansatz, using a classical optimization procedure to update the parameters. 

The Ansatz in a VQA can be either problem-agnostic~\cite{kandala2017hardware} or retain the structure of the target problem~\cite{barkoutsos2018quantum}. Problem-agnostic Ansatze are typically designed to minimize the effect of noise by accounting for hardware constraints. However, they come at the cost of limited expressivity compared to Ansatze that fully encode the problem structure. 
Designing expressive and shallow Ansatze remains an open question~\cite{holmes2022connecting}. 

Here, we focus on VQAs to solve combinatorial optimization problems, which are prevalent across industry applications such as communication~\cite{peterson2007computer}, finance~\cite{markowitz2008portfolio, egger2020quantum}, and vehicle routing~\cite{domino2022quadratic, pascariu2024formulation, sbihi2010combinatorial}. Many relevant combinatorial optimization problems are hard to solve exactly~\cite{karp1975computational}. In addition, for some of them, polynomial-time approximate algorithms that reach an arbitrarily accurate approximation of the optimal solution are not available~\cite{zuckerman1996unapproximable}. 
Therefore, many classical algorithms designed to solve such problems are heuristics. Crucially, they work well in practice, thus motivating a similar heuristic approach for quantum computing~\cite{abbas2024challenges}.

A foundational VQA for combinatorial optimization problems is the Quantum Approximate Optimization Algorithm (QAOA)~\cite{farhi2014quantum, blekos2024review}. Its trotterized Ansatz structure alternates $p$ time evolutions of a Hamiltonian representing the target problem and of a problem-independent Hamiltonian. The structure is inspired from the adiabatic theorem in physics, and thus inherits its convergence guarantee in the limit $p\to\infty$~\cite{finnila1994quantum, farhi2014quantum, rajak2023quantum}.  

Finite-depth QAOA is a heuristic algorithm, which implies that its performance must be thoroughly investigated and benchmarked on a case-by-case basis~\cite{abbas2024challenges, koch2025decathlon}. In particular, research often investigates QAOA for quadratic unconstrained binary optimization problems (QUBOs), such as Max-Cut, to benchmark hardware performance~\cite{zhou2020quantum, brandhofer2022benchmarking, willsch2020benchmarking, santra2024squeezing} and validate new algorithms~\cite{tate2023warm, egger2021warm}. However, Max-Cut is easy to solve classically at low graph densities and up to 10\,000 nodes~\cite{rehfeldt2023faster}. Solving it using QAOA would only yield approximate solutions even in an ideal case~\cite{gamarnik2021overlap, lykov2023fast, lykov2023sampling, akshay2020reachability}, thus making it a poor candidate to deliver a quantum advantage. Hence, recent studies investigated high-order unconstrained binary optimization problems (HUBOs) with cubic and quartic terms. Certain HUBOs can be very hard to solve, even at a low number of decision variables~\cite{romero2024bias}. For instance, the largest exactly solved instance of the Low Autocorrelation Binary Sequence (LABS) problem has only $66$ decision variables~\cite{brest2021low, packebusch2016low}. Importantly, Ref.~\cite{shaydulin2024evidence} suggests a scaling advantage for QAOA over Gurobi~\cite{gurobi} or CPLEX~\cite{cplex} in solving LABS. 

In this work, we first show in Section~\ref{sec:2} how the quantum resources requirements of QAOA scale unfavorably for HUBOs by considering two test cases: a fully-connected four-local Hamiltonian, and the LABS problem. Next, we introduce in Section~\ref{sec:3} two approximate quadratization procedures to reduce the QAOA circuit depth without qubit overhead. Motivated by the idea of approximating the circuit Ansatz associated with a given problem, we benchmark how simplifications in the QAOA Ansatz improve measured approximation ratios when executing Max-Cut on hardware in Section~\ref{sec:4}. Finally, in Section~\ref{sec:5}, we argue for the need of a more mathematically motivated quadratization and suggest some ideas in this direction.

\section{Executing QAOA on digital quantum computers}\label{sec:2}

We consider combinatorial optimization problems defined over $n$ binary variables $x \in \{ 0,1\}^n$ which minimize an objective function $f(x)$. Quantum computers can tackle such problems by mapping $f(x)$ to an Ising Hamiltonian $H_C$ whose ground state minimizes $f(x)$. This is done through the change of variables $x_i=\frac{1-z_i}{2}$ and promoting $z_i$ to Pauli $Z$ operators $Z_i$~\cite{lucas2014ising}. Then, we use QAOA to find the ground state of $H_C$ by sampling from an Ansatz
\begin{equation}
    \ket{\psi(\bm{\beta}, \bm{\gamma})} = \prod_{q=1}^p e^{-i \beta_q H_M} e^{-i \gamma_q H_C}\ket{+}^{\otimes n}
\end{equation}
with optimized variational parameters $(\bm{\beta}, \bm{\gamma})$.
Here, we consider a problem-independent mixer operator \mbox{$H_M = -\sum_{i=0}^{n-1} X_i$}, although other mixing Hamiltonians have been developed to enforce constraints or leverage warm-starts~\cite{maciejewski2024design, zhu2022adaptive, fuchs2022constraint, egger2021warm, he2023alignment, niu2019optimizing}. 

\figureOne

\setlength{\tabcolsep}{8.8pt} 
\begin{table*}[htbp]
\centering
\begin{tabularx}{\textwidth}{c  c  c  c  c  c  c  c  c } 
 \hline
 \hline
  Problem &  \multicolumn{2}{c}{$4^{\text{th}}$ order terms}  & \multicolumn{2}{c}{$2^{\text{nd}}$ order terms}&  \multicolumn{2}{c}{All-to-all connectivity} & \multicolumn{2}{c}{Linear connectivity} \\[0.3ex] 
   & scaling & $n=16$ & scaling & $n=16$ & \# $CZ$ gates  & $CZ$ gate depth & \# $CZ$ gates & $CZ$ gate depth\\ [0.5ex] 
 \hline
 $H_2^{\text{full}}$ & -  & - & $O(n^2)$ & 120 & 240 & 32 & 507& 103\\ [0.3ex] 
 LABS & $O(n^3)$ & 252 & $O(n^2)$ & 56 & 1218 & 1071 & 4685 & 2978\\ [0.3ex] 
 $H_4^{\text{full}}$ & $O(n^4)$ & 1820 & -  & - & 4732 & 4622 & 17813 & 13920\\ [0.3ex] 
 \hline
 \hline
\end{tabularx}
\caption{Complexity scaling of selected problems. The trends in the first and third columns are shown in Fig.~\ref{fig:hardness}. The numerical values in the remaining columns correspond to problem instances with 16 variables. }
\label{table1}
\end{table*}

QUBOs are unconstrained combinatorial optimization problems with a polynomial $f(x)$ of degree at most two. As such, their cost Hamiltonian is reads as
\begin{equation}
    H_{\text{QUBO}} = \sum_{i<j} w_{ij} Z_iZ_j +\sum_j w_j Z_j.
\end{equation}
The main challenge associated with executing QAOA on quantum processors is implementing  the time evolution of the cost Hamiltonian. In particular, the more terms $H_C$ contains, the denser and deeper the resulting circuits become, which results in noisier samples. The maximum number of gates in the time-evolution circuit of a fully-connected QUBO $H_2^{\text{full}}$ scales as $O(n^2)$, with a $O(n)$ two-qubit gate depth\footnote{We only consider two-qubit gates in gate counts as these are the main source of errors. For instance, the median error rates of the $CZ$ gates is $8.30 \cdot 10^{-3}$, compared to $2.23 \cdot 10^{-4}$ for single-qubit gates on \emph{ibm\_fez}. Similarly, when computing circuit depth, we refer only to the depth required to implement two-qubit gates.}, see the purple lines in Fig.~\ref{fig:hardness}. When transpiled to a line of qubits the complexity exponent remains unchanged, but the scaling prefactor increases by 1.5 due to the additional SWAP gates required to reach all-to-all connectivity, see details in App.~\ref{sec:scaling_trends}.

Furthermore, the locality of the gates in $\exp(-i\gamma H_C)$ significantly impacts the circuit depth. A gate $\exp(-i\gamma Z_iZ_j)$ is implemented by a single $R_{ZZ}$ rotation if the qubits $i$ and $j$ are connected on the hardware. Otherwise, additional SWAP gates are needed to bring qubits $i$ and $j$ in contact. Minimizing the SWAP overhead is desirable since each SWAP gate is built from three noisy entangling gates. Implementing $\exp(-i\gamma H_C)$ is even more challenging for HUBOs, whose terms require more stringent connectivity constraints, i.e. implementing $\exp(-i\gamma Z_iZ_jZ_kZ_l)$ requires bringing close together qubits $(i,j,k,l)$. 
Therefore, HUBOs are more complicated to implement than QUBOs as they require implementing possibly many non-local gates. We now investigate the gate count and depth for two examples of HUBOs, i.e. the LABS problem and a fully-connected fourth-order Hamiltonian $H_4^{\text{full}}$. 

The LABS problem has applications in signal-processing~\cite{golay977LABS} and statistical physics~\cite{bernasconi1987title}. It aims to minimize the sidelobe energy 
\begin{equation}
    \mathcal{E}_{\text{sidelobe}} = \sum_{k=1}^{n-1} C_k(z)^2
\end{equation}
of a length $n$ spin sequence $z\in\{-1, 1\}^n$. Here, the autocorrelation functions are
\begin{equation}
    C_k(z) = \sum_{i=1}^{n-k} z_i z_{i+k}.
\end{equation}
Expanding the sums in the expression of the sidelobe energy yields the Hamiltonian~\cite{shaydulin2024evidence}
\begin{align} \label{eq:LABS_Ham}
    H_{\text{LABS}} =& 2 \sum_{i=1}^{n-3} Z_i \sum_{t=1}^{\lfloor \frac{n-i-1}{2} \rfloor } \sum_{k = t+1}^{n-i-t} Z_{i+t} Z_{i+k} Z_{i+k+t} \notag \\
    & + \sum_{i=1}^{n-2} Z_i \sum_{k=1}^{\lfloor \frac{n-i}{2} \rfloor} Z_{i+2k}.
\end{align}
Solving LABS with QAOA requires implementing the time-evolution operator associated with $O(n^3)$ four-local terms and $O(n^2)$ two-local terms. The dominant $O(n^3)$ scaling is shown by the thicker blue lines in Fig.~\ref{fig:hardness}. For example, a LABS instance with 16 decision variables has 252 four-local terms and 56 two-local terms. This requires $252 \times 6 + 56 \times 2  = 1624$ two-qubits gates assuming an all-to-all connectivity, where the constant factors 6 and 2 are due to the two-qubit gate decomposition, see Figs.~\ref{fig:main_ideas}(a,b). After transpilation using Qiskit with the highest optimization level 3, this number reduces to 1218 through cancellation of neighboring two-qubit gates~\cite{javadi2024quantum}. However, when compiled to linearly connected qubits, the two-qubit gate count increases to 4685 with a two-qubit gate depth of 2978, see the thin blue line in Fig.~\ref{fig:hardness}. Implementing such deep circuits on current noisy hardware is a formidable challenge. For instance, on \emph{ibm\_fez}, the median $T_1$ time of $146~{\rm \mu s}$ fits approximately only 1738 layers of two-qubit gates with a median gate time of $84~{\rm ns}$. 

The hardest fourth order problem to implement is a fully-connected four-local Hamiltonian $H_4^{\text{full}}$, which has a similar structure as the Sachdev-Kitaev-Ye model of Majorana fermions~\cite{rosenhaus2019introduction, luo2019quantum, kobrin2021many}, but which is expressed in terms of Pauli operators instead of fermionic ones. Concretely, we define

\begin{equation}\label{eq:SYK_Ham}
    H_4^{\text{full}} = \sum_{i<j<k<l}^{{n \choose 4}} J_{ijkl} Z_{i}Z_{j}Z_{k}Z_{l},
\end{equation}
$J_{ijkl}$ being real coefficients. 
The resulting circuits have $O(n^4)$ two-qubit gates, and $O(n^3)$ two-qubit gate depth, see teal line in Fig.~\ref{fig:hardness}. Implementing QAOA for $H_4^{\text{full}}$ is thus more challenging than for LABS, and considerably more so than for a fully-connected QUBO. This is true not only in the asymptotic limit, but also for instances with as few as 16 variables which require 4732 $CZ$ gates and a depth of $4622$ assuming all-to-all connectivity, and more than three times as much for a linear qubit topology, see Tab.~\ref{table1}.

Finally, these deep circuits result in low sampling rates on hardware. We estimate the rate at which we generate samples by the inverse of the duration of the quantum circuit. Furthermore, we lower bound the duration of a depth-one QAOA with the duration of running $e^{-i \gamma H_C}$, which is the product of the two-qubit gate duration $\bar{T}$ and the two-qubit gate depth. Extrapolating the data in Fig.~\ref{fig:hardness} to a LABS system with 127 qubits, corresponding to one of the largest lines of superconducting qubits available~\cite{javadi2024quantum}, we conclude that it can be implemented with a two-qubit gate depth of about $8.82\cdot 10^6$. Taking $\bar{T} = 84~{\rm ns}$, corresponding to the median $CZ$ gate duration of \emph{ibm\_fez}, this results in a runtime of $0.741~{\rm s}$ per sample. Therefore, we can generate samples at a rate of at most $1.35~{\rm s}^{-1}$, which is six orders of magnitude slower than the best-known classical methods for the same problem size~\cite{brest2018heuristic}. 
 
\section{Approximate quadratization} \label{sec:3}

Compiling quartic HUBO problems to hardware-native instructions results in very deep circuits, while QUBOs are optimally compiled in linear depth with SWAP networks~\cite{maciejewski2024design, matsuo2023sat}. 
This motivates us to generate the ansatz based on QUBOs defined in terms of the target HUBO. 
For example, an exact quadratization that exactly conserves the full structure of the HUBO replaces each four-local term with one constant, four linear, and six quadratic terms at the expense of adding two ancilla qubits per four-local term~\cite{mandal2020compressed}. 
This implies that implementing four-local Hamiltonians with $O(n^3)$ terms, such as LABS, or $O(n^4)$ terms, such as $H_4^{\text{full}}$, requires circuits with a qubit overhead scaling as $O(n^3)$ or $O(n^4)$, respectively. 
This is impractical due to the limited qubit count of the current hardware. We therefore explore \emph{approximate} quadratization methods to project a generic HUBO onto a QUBO without adding decision variables.

\subsection{Methods} \label{sec:circuit_design}

We study two methods to quadratize HUBOs.
The first is based on a hypergraph clique expansion, and the second is based on a fully-connected quadratic Hamiltonian whose gate weights are treated as additional variational parameters.
The price paid to keep the qubit count constant is an approximation of the original problem. 

In the hypergraph clique expansion~\cite{suppakitpaisarn2024utilizing} we treat a high-order Hamiltonian $H_C=\sum_{i=0}^{K-1}\omega_iP_i$ as a hypergraph $\mathcal{G}=(\mathcal{E}, V)$.
Here, $\omega_i$ is the coefficient of the Pauli $P_i$.
We associate each Pauli term $P_i=Z_{i_1}Z_{i_2}...Z_{i_k}$ with a hyperedge $e = (i_1, i_2, ..., i_k)\in\mathcal{E}$.
To quadratize $H_C$, we create a weighted graph $G=(E,V)$, resulting in a quadratic Hamiltonian $H_2=\sum_{ij}\omega_{ij}Z_iZ_j$.
We add an edge to $E$ for each pair of nodes $(i,j)$ within a hyperedge $e$, as shown in Fig.~\ref{fig:main_ideas}(c).
Therefore, each hyperedge $e$ with $k$ nodes in $\mathcal{E}$ generates a \textit{$k$}-clique in $G$ with edge-set $E(e)$.
Since a given pair of nodes $(i,j)$ might belong to multiple hyperedges of $\mathcal{E}$, the weight $w_{ij}^\text{opt}$ we assign to it minimizes the $l_2$-norm, i.e. the sum of squared differences, with respect to the weights $w(e)$ of all hyperedges $e$ that contain $(i,j)$. The $l_2$-norm is chosen according to the quadratization given in the literature~\cite{suppakitpaisarn2024utilizing}. 
In particular, we choose $\omega_{ij}$ to minimize

\begin{align}\label{eq:clique_proj}
f_2(\boldsymbol{\omega})=\sum_{e\in\mathcal{E}}\sum_{(i,j)\in E(e)}\alpha_e[\omega_{ij}-\omega_e]^2. 
\end{align}
Here, $\omega_e$ is the weight of hyperedge $e$.
The weight $\alpha_e=\omega_e/\sum_{e\in\mathcal{E}}\omega_e$ prevents low-weight edges from having a disproportionate impact on the quadratization.
We use the resulting quadratic Hamiltonian
\begin{equation}
H_2=\sum_{(i,j) \in E} \omega_{ij}^{\text{opt}}Z_iZ_j
\end{equation}
to generate the Ansatz circuit of QAOA, while still evaluating the objective function with respect to the original quartic Hamiltonian. Concretely, after the circuit prepares a state 
\begin{equation}\label{eq:psiQUBO}
    \ket{\psi_2 (\bm{\beta}, \bm{\gamma})} = \prod_{q=1}^p e^{-i\beta_q H_M}e^{-i\gamma_q H_2}\ket{+}^{\otimes n},
\end{equation}
we minimize the energy $\langle \psi_2(\bm{\beta}, \bm{\gamma})|H_C| \psi_2(\bm{\beta}, \bm{\gamma})\rangle$.\\

\figureTwo

The second method consists of generating an Ansatz circuit with a potentially fully quadratic Hamiltonian whose parameters $\boldsymbol{\theta}$ are simultaneously optimized with $(\boldsymbol{\beta}, \boldsymbol{\gamma})$ to minimize the energy of $H_C$.
In particular, we perform the joint minimization
\begin{align}\label{eqn:qaoa_proj}
    \min_{\boldsymbol{\theta}, \boldsymbol{\beta}, \boldsymbol{\gamma}}\braket{\psi'_2(\boldsymbol{\theta}, \boldsymbol{\beta}, \boldsymbol{\gamma})|H_C|\psi'_2(\boldsymbol{\theta}, \boldsymbol{\beta}, \boldsymbol{\gamma})},
\end{align}
where the Ansatz
\begin{align}
    \ket{\psi'_2(\boldsymbol{\theta}, \boldsymbol{\beta}, \boldsymbol{\gamma})}=\prod_{q=1}^pe^{-i\beta_q H_M}e^{-i\gamma_qH_2'(\boldsymbol{\theta})}\ket{+}^{\otimes n}
\end{align}
is generated by the Hamiltonian
\begin{align}
    H_2'(\boldsymbol{\theta})=\sum_{i>j}\theta_{ij}Z_iZ_j+\sum_{i=0}^{n-1}\theta_iZ_i.
\end{align}
The parameters $\boldsymbol{\theta}=\{\theta_{ij}, \theta_i\}$ are optimized at the same time as the QAOA angles $\boldsymbol{\beta}$ and $\boldsymbol{\gamma}$.
This differs from other Ans\"atze such as the multi-angle Ansatz (ma-QAOA)~\cite{herrman2021multiangle}.
In ma-QAOA the Hamiltonian generating the Ansatz is built from the exact same terms as the cost operator $H_C$ but each Pauli term $k$ is associated with an independent QAOA angle $\gamma_{q,k}$ at each QAOA layer.
Therefore, our $H_2'(\boldsymbol{\theta})$ defines a quadratization of the original quartic Hamiltonian $H_C$.

These quadratizations introduce an inconsistency between the cost operator and the operator that creates the Ansatz. 
Therefore, the convergence guarantee of a typical QAOA in the limit $p\to\infty$ is lost. 
Instead, the approach is inherently heuristic as it is based on an Ansatz circuit that retains only some of the structure of $H_C$, but which considerably reduces the circuit depth.
The exact depth reduction depends on the density of the original problem. 
In the case of $H_2'(\boldsymbol{\theta})$ the convergence as $p\to\infty$ may still persist as long as $H_2'(\boldsymbol{\theta})$ shares the same ground state as $H_C$.
The largest gate count reduction occurs for dense Hamiltonians, such as $H_4^\text{full}$, for which the reduction scales as $O(n^2)$, see Fig.~\ref{fig:hardness}. 
For LABS, which has $O(n^3)$ terms, the reduction factor is $O(n)$.

\subsection{Expected performance of the quadratizations}

We now benchmark the accuracy of the quadratizations by investigating depth-one and depth-two QAOA.
We perform this assessment on the LABS Hamiltonians with 12 decision variables. 

The depth-one QAOA angles $(\beta_1^\star, \gamma_1^\star)$ are trained with a coarse 2D grid scan with 15 points in the interval $[0,\pi/2]$ followed by a Scipy optimization with COBYLA to refine the QAOA angles of the grid search.
The depth-two angles are found with COBYLA from the transition state $(\beta_1^\star, 0, \gamma_1^\star, 0)$ as initial point~\cite{sack2023recursive}.
The parameters $\boldsymbol{\theta}$ in $H_2'(\boldsymbol{\theta})$ and the depth-$p$ QAOA angles are trained simultaneously with Scipy minimize and COBYLA to minimize the energy of $H_C$, see Fig.~\ref{fig:projection_convergence}.
The parameters in the clique expansion are optimized with Scipy and COBYLA independently from the QAOA parameters and converge to a low value of $f_2(\boldsymbol{\omega})$, see Fig.~\ref{fig:projection_convergence}.

The optimized quantum circuits are sampled with the state-vector simulator of Qiskit.
These samples are then converted to a cumulative distribution function of the energy of the LABS Hamiltonian $H_C$.
We compare the clique expansion, the $H_2'(\boldsymbol{\theta})$ projection, standard QAOA, and uniform sampling.
The performance of the clique expansion is very close to random samples, see Fig.~\ref{fig:projection_cdf}.
Both standard depth-one and two QAOA outperform random sampling and the clique expansion, see dashed curves in Fig.~\ref{fig:projection_cdf}.
The depth-two standard QAOA, for example, samples the optimal -28 energy bit-string with a probability of 0.8\%.
The projection $H_2'(\boldsymbol{\theta})$ does not find the optimal bit-string but produces good, yet sub-optimal, solution with high probability.
For example, the depth-two QAOA with the $H_2'(\boldsymbol{\theta})$ projection produces samples with a 97.1\% approximation ratio, defined by $(E(x)-E_\text{max})/(E_\text{min}-E_\text{max})$, with a probability of 95.7\%.
Here, $E(x)$, $E_\text{min}$, and $E_\text{max}$ are the LABS energies of the candidate solution $x$, the lowest, and highest energies, respectively.
The quadratization yields an interesting tradeoff since it only approximately solves the 12 node LABS problem but with a significantly shorter circuit depth.
Indeed, the two-qubit gate depth and count of the Ansatz circuits based on the quadratization are 68 and 374 when transpiled with a SWAP network on a line of qubits~\cite{weidenfeller2202scaling}.
By contrast, transpiling the standard QAOA Ansatz with the Qiskit transpiler at optimization level three on a fully connected network of qubits creates circuits with a two-qubit gate count and depth of 600 and 417, respectively.
The increased gate count will likely worsen when the circuit is transpiled on a line of qubits.
Irrespective of the qubit topology, this example shows the saving in quantum computational resources that the quadratization yields.

\figureThree

\figureFour

The clique-based quadratization fails to produce good samples because it does not preserve the ground state of the problem.
For example, the clique quadratization of the quartic term $Z_3Z_2Z_1Z_0$ produces the fully connected Hamiltonian $\sum_{i<j}^3Z_iZ_j$.
The ground state of $Z_3Z_2Z_1Z_0$ is made of bit-strings with Hamming weight 1 or 3.
By contrast, the ground state of the fully-connected quadratic Hamiltonian have Hamming weight 2.
This also helps explain the success of the quadratization in Eq.~(\ref{eqn:qaoa_proj}) which constructs an Ansatz that minimizes the energy of the cost Hamiltonian.

\subsection{Noisy simulations} \label{sec:noisy_simulations}

\figureFive

We now study the performance of the $H_2'(\boldsymbol{\theta})$ quadratization and standard depth-two QAOA by setting up a noisy simulation in which we apply a depolarizing channel with strength $\lambda$ after each two-qubit gate. In Fig.~\ref{fig:simulations}, we show how the energy averaged over the best fraction $\alpha\in \{0.01, 0.05. 0.1 \}$ of all samples evolves with $\lambda$. By varying $\alpha$ we see how the noise impacts the tail of the distribution of samples produced by the quantum circuit.

The quadratized Ansatz is less sensitive to noise, making it a viable option for noisy quantum computers. Generally, as $\lambda$ increases, the average energies of the best $\alpha$-fraction of samples increase, corresponding to a worse performance, as expected. Importantly, at $\alpha=0.01$ the Ansatz based on the QUBO projection (purple) produces samples that are statistically insensitive to the noise. By contrast, the energy of the QAOA Ansatze on a line of qubits (blue) and all-to-all connected qubits (pink) show a strong noise dependence, see Fig.~\ref{fig:simulations}(a).
At $\alpha = 0.1$, the QUBO Ansatz is $25\%$ and $36\%$ less sensitive\footnote{Noise sensitivity is computed as $$ \frac{ \Big\langle E_{\text{QUBO}}(\lambda = 0.01) \Big\rangle - \Big\langle E_{\text{QUBO}}(\lambda = 0.001) \Big\rangle  }{  \Big\langle E_{\text{LABS}}(\lambda = 0.01) \Big\rangle - \Big\langle E_{\text{LABS}}(\lambda = 0.001) \Big\rangle }$$} than the other two types of Ansatze, respectively, see Fig.~\ref{fig:simulations}(c). These results indicate that the samples produced from a quadratized Ansatz are less sensitive to noise at the scale and noise strengths considered here.

The effectiveness of quadratizing the Ansatz depends on the noise strength and set $\alpha$ values. At $\alpha =0.01$ and noise strength $\lambda = 0.001$, the standard QAOA Ansatz outperforms the quadratization. However, this advantage vanishes at $\lambda = 0.004$ on a line connectivity. When considering the $5\%$ best samples, the quadratization is better than the standard QAOA for both all-to-all and line connectivity, regardless of noise strength, see Fig.~\ref{fig:simulations}(b). 

These data show the benefits of quadratizing the Ansatz when working with hardware with small to intermediate noise relative to the size of the optimization problem. Furthermore, they reinforce the trade-off discussed in Fig.~\ref{fig:projection_cdf}, namely that the quadratized Ansatz can increase the frequency of sampling low-energy, yet sub-optimal, states.

\section{Hardware-native circuits for QUBOs} \label{sec:4}
\figureSix

We now study an approximate hardware-friendly compilation method of Ansatz circuits for QUBOs. The standard approach to route an arbitrary QAOA circuit utilizes a predetermined network of SWAP gates tailored to the qubit connectivity~\cite{weidenfeller2202scaling, matsuo2023sat, maciejewski2024design, Harrigan_2021}. We use a line of qubits to benefit from two-qubit gate cancellations, even though the IBM quantum hardware connectivity has a heavy-hex structure~\cite{weidenfeller2202scaling}. To fully implement $e^{-i\gamma H_C}$, we apply $k_\text{max}$ alternating layers of even and odd SWAP layers\footnote{A SWAP layer is a depth-one circuit which applies SWAP gates to pairs of neighboring qubits, and it is called even if the first pair is $(0,1)$ and odd if it is $(1,2)$.}, see Fig.~\ref{fig:main_ideas}(d). 
Full connectivity is reached after $n-2$ SWAP layers.
The SWAP network thus allows us to derive an approximate Ansatz by including only $k<k_\text{max}$ layers when implementing $e^{-i\gamma H_C}$. This allows us to implement only the time evolution of a subset $H_C(k)$ of the terms in $H_C$. 
The truncated $H_C(k)$ is based on the subgraph $G_k = (E_k, V) \subseteq G = (E,V)$. Details on the transpilation are given in App.~\ref{sec:transpilation}. Importantly, the size $|E_k|$ depends on the initial mapping of the decision variables to the qubits. 
Following Ref.~\cite{matsuo2023sat} we optimize this initial mapping with a SAT solver to minimize the number of SWAP layers needed to implement the cost operator.
Once the Ansatz is created, we optimize the QAOA parameters by minimizing the energy 
\begin{align}
\min_{{\bm \beta}_k,{\bm \gamma_k}}\langle \psi_k({\bm \beta}_k, {\bm \gamma}_k)|H_C|\psi_k({\bm \beta}_k,{\bm \gamma_k})\rangle
\end{align}
where
\begin{align}
    \ket{\psi_k({\bm \beta}_k,{\bm \gamma_k})}=\prod_{j=1}^pe^{-i\beta_{j,k}H_M}e^{-i\gamma_{j,k}H_C(k)}\ket{+}^{\otimes n}.
\end{align}

Our approach is similar to the time-block Ansatz design presented in Ref.~\cite{maciejewski2024design}. The main distinction is that a time-block Ansatz assigns different, independently optimized parameters to the edges of $H_C(k)$. Concretely, they replace the values $\gamma_{i,j}$ with corresponding vectors $\{\gamma_{i,j}^{e}, \forall e \in E\}$.

We benchmark the truncated Ansatz on Max-Cut problems~\cite{karp1975computational} for all $k \in [0, k_{\max}]$. We thus aim to split the set $V$ of $n$ nodes of a graph $G = (E,V)$ in two parts such that the sum of the edges connecting the two subsets is maximized. Formally, a variable $x_i$ is assigned to each node $i \in V$ to indicate which side of the cut the node is on. Solving Max-Cut is then equivalent to finding the binary string $x$ that maximizes 
\begin{equation}\label{eq:cost_function}
    f(x) = \sum_{(i,j) \in E} \Big( x_i (1-x_j) + x_j (1-x_i) \Big).
\end{equation}
Mapping $f(x)$ to a Hamiltonian~\cite{lucas2014ising} results in the cost operator
\begin{equation}
    H_{\text{Max-Cut}} = \frac{1}{4} \sum_{(i,j) \in E}  Z_i Z_j,
\end{equation}
with a minimum energy eigenstate that maps to the maximum cut. We measure the quality of a candidate solution $x$ with the approximation ratio
\begin{equation}
    r = \frac{f(x)}{\max_x f(x)} \equiv \frac{E_{\max} - E(x)}{E_{\max} - E_{\min}} \in [0,1],
\end{equation}
where 
$E(x) = \langle x |H_{\text{Max-Cut}} | x \rangle$ is the energy of the basis state $\ket{x}$. The objective value and the energy are related by $2f(x) = |E| - E(x)$. The ground state of $H_\text{Max-Cut}$ has $r=1$. On average, uniformly sampled candidate solutions have $r=0.5$. The best known polynomial-time randomized rounding algorithm, the Goemans-Williamson algorithm~\cite{goemans1995improved}, produces samples with $r\simeq 0.878$, on average. For $s$ shots, an average $r$ is computed over the values $r_k$ computed from corresponding bitstrings $x_k$.  

We generate three random three-regular (RR3) graphs with 40 nodes. 
Each graph can be SAT mapped following Ref.~\cite{matsuo2023sat} in $k_\text{max}=9$ SWAP layers. Next, we generate an Ansatz circuit with $k\in[0, k_\text{max}]$ SWAP layers and depth $p\in \{1,2,3\}$. We train the parameters in each Ansatz using an MPS-based simulator to evaluate the energy with a bond dimension of 20, see App.~\ref{sec:RR3_parameters}. 
We observe a monotonic increase with $k$ of the approximation ratio evaluated with the MPS, see Fig.~\ref{fig:rr3}(a).
This shows that the QAOA performance improves as the Ansatz retains more of the problem structure.
The circuits are then executed on \emph{ibm\_fez} from which we draw 50\,000 samples.
The samples at $k=9$ are better than those measured by Sack and Egger on an Eagle device~\cite{sack2024blackline}, see the black dashed line in Fig.~\ref{fig:rr3}(b). 
This is expected since \emph{ibm\_fez} is a Heron device with lower error rates than Eagle devices. 
Crucially, this data shows significant gains in solution quality when the truncated Ansatz is applied.
In the presence of hardware noise, there is an optimal point $(k,p)$, which represents the maximum depth beyond which increasing either $k$ or $p$ reduces $r$, see Fig.~\ref{fig:rr3}(b).
On \emph{ibm\_fez}, we find this optimal point at depth-three with two SWAP layers.
Once noise begins to reduce the approximation ratio, the higher depth QAOA degrades faster than the lower depth QAOA.

Increasing $k$ and $p$ increases the differences between the ideal and the hardware results, compare Fig.~\ref{fig:rr3}(a) and (b). 
We can recover a noise-free approximation ratio with the ${\rm CVaR}_\alpha$ aggregation function~\cite{barkoutsos2020cvar, barron2024cvar}, which keeps only an $\alpha\in[0,1]$ fraction of the best samples $x$ measured by $f(x)$.
Here, we fit $\alpha$ such that the approximation ratio measured with ${\rm CVaR}_\alpha$ matches the noiseless one obtained with the MPS simulator.
The optimization of the $\alpha$ parameter is carried out with COBYLA and results in a root mean squared error of $10^{-6}$ between the 90 MPS approximation ratios and the 90 ${\rm CVaR}$ ones (three graphs, three depths and ten values of $k$).
As expected, as $k$ and $p$ increase, we discard more shots to recover a noiseless expectation value, see solid lines in Fig.~\ref{fig:rr3}(c).
This noise strength is then compared to a theoretical amount of noise based on error rates reported by \emph{ibm\_fez}.
To this end, we compute $\alpha_\text{th}=1/\sqrt{\gamma}$, where $\gamma$ measures the noise strength of the circuit.
We make the simple assumption that 
\begin{align}\gamma=\gamma_0^{p(2+3\lceil\frac{k}{2}\rceil)}\gamma_1^{p(2+3\lceil\frac{k}{2}\rceil-3)}.
\end{align}
Here, $\gamma_0$ and $\gamma_1$ represent the layer fidelity~\cite{mckay2023benchmarking} of the even and odd layer of $CZ$ gates, respectively.
The factors $\gamma_i^{2p}$ account for the first hardware-native layers of $R_{ZZ}$ gates which transpile to four layers of $CZ$ gates per QAOA layer~$p$ on \emph{ibm\_fez}.
The terms $\smash\gamma_0^{3p\lceil\frac{k}{2}\rceil}$ and $\smash\gamma_1^{3p\lceil\frac{k}{2}\rceil-3p}$ account for the even and odd SWAP layers, respectively, which transpile into three layers of $CZ$ gates each.
We approximately express $\gamma_i$, where $i \in \{0,1 \}$, using the errors $\epsilon_{j,j+1}$ of the native entangling gates between the pairs of qubits $j$ and $j+1$ reported for the used quantum processor.
For the even and odd layers of two-qubit gates, we have
\begin{align}
    \gamma_i=\prod_{j=0}^{\lfloor \frac{n-1}{2} \rfloor -1}(1+\epsilon_{2j+i,2j+i+1}),
\end{align}
where the indexing is based on an appropriate indexing of the two qubit gates and we assume $n \geq 3$.
This simple $1/\sqrt{\gamma}$ estimation of $\alpha$ already produces excellent qualitative agreement with the fitted $\alpha$, the solid lines in Fig.~\ref{fig:rr3}(c) show the same trend as the dashed ones.
At a low number of SWAP layers, we overestimate the sampling overhead since the transpiler removes the unnecessary SWAP gates that the SWAP strategy added.
When the circuits become deep, e.g. $k=9$ and $p=3$, we tend to underestimate the sampling overhead since we do not account for single-qubit gate noise.
In summary, we can (i) trade approximation ratio for a partial implementation of the problem with less noise and (ii) recover a noiseless setting by post-selecting and drawing more samples.
This sampling overhead scales exponentially in the circuit depth.

\section{Discussion and Conclusions} \label{sec:5}

Using noisy quantum hardware to solve non-hardware-native dense optimization problems remains a challenge, which high-order optimization problems exacerbate. 
For dense four-local Hamiltonians, the circuit gate count scales as $O(n^4)$. The circuit depth limits the size of the LABS problem to about 30 decision variables before the noisy hardware samples are effectively random, for a depth-one QAOA and a depolarizing noise strength of $\lambda = 0.001$. This threshold is expected to be even lower for denser problems than LABS. 

To reduce the circuit depth without increasing its width we explore approximate quadratizations of the HUBO problem. These constructions sacrifice the performance guarantee of QAOA in the $p \to \infty$ limit, but recover an Ansatz with a depth linear in problem size~$n$.
The clique expansion method, originally designed for mathematical applications~\cite{suppakitpaisarn2024utilizing}, does not guarantee that the quadratized Hamiltonian preserves the ground state of the original problem. 
Indeed, an ideal quatratization would ensure that the ground state of the quadratized Hamiltonian and the quartic one coincide.
By contrast, optimizing the parameters of a quadratized Hamiltonian Ansatz partially overcomes this issue. The resulting samples have a high probability to have an approximation ratio close to one while being sub-optimal.

Our simulations confirm that the standard QAOA Ansatz cannot be used in utility-scale applications because it inevitably requires error correction. Interestingly, they also show that a quadratized Ansatz yields better samples than the standard QAOA Ansatz once a threshold noise strength is reached.
This makes the quadratized Ansatz particularly appealing for current quantum hardware. 

Future work may further investigate on which problems this quadratization approach may be maximally efficient. Furthermore, our approach can be extended by developing hybrid methods that mix our approximate quadratization with an exact implementation of a subset of the high-order terms.  
In addition, one could choose to exactly quadratize the terms in a subset $E'$ of the fourth order terms as in \cite{mandal2020compressed}, which would come at a cost of $2|E'|$ qubit overhead.

The second contribution of this paper is designing hardware-friendly Ansatze for QUBOs. We use SWAP Strategies \cite{weidenfeller2202scaling, matsuo2023sat, Harrigan_2021, maciejewski2024design} to create a truncated cost Hamiltonian that approximates the target one.
The resulting approximation ratio monotonically increases with the number of SWAP layers $k$ in a noiseless setting.
However, in the presence of noise, there is an optimal pair $(k,p)$ that maximizes the average approximation ratio. In future work, it is worth investigating whether using SWAP Strategies improves the results found in Ref.~\cite{maciejewski2024design}.

Further, optimizing the QAOA variational parameters remains a challenge due to vanishing gradients~\cite{larocca2024review, holmes2022connecting, arrasmith2022equivalence, mcclean2018barren}, and an abundance of local minima~\cite{bittel2021training}. Thus, solving dense problems with the typical QAOA Ansatz requires significant quantum and classical resources. Progressively building-up the Ansatz and optimizing its parameters one SWAP layer at a time may mitigate this effect. Future work will investigate the effect of additionally optimizing the weights of the truncated QUBO, similarly to our HUBOs quadratization scheme. 

In summary, our work highlights the challenges in implementing high-order optimization problems on quantum hardware. We show how these challenges can be mitigated by carefully designing the Ansatz circuit from which to sample. 
\section{Acknowledgments}

The authors acknowledge Stefan Woerner for stimulating discussions.
D.J.E.~acknowledges funding within the HPQC project by the Austrian Research Promotion Agency (FFG, project number 897481) supported by the European Union – NextGenerationEU. S.D.~acknowledges funding from NCCR SwissMAP and the ETH Zurich Quantum Center. 

\appendix 
\numberwithin{equation}{section}

\section{Circuit complexity scaling}\label{sec:scaling_trends} 

We now give some intuition on the number of entangling gates discussed in Section~\ref{sec:2} and depicted in Fig.~\ref{fig:hardness}(a). Assuming all-to-all qubit connectivity, implementing all \mbox{$n(n-1)/2$} $R_{ZZ}$ terms requires \mbox{$n(n-1)$} entangling two-qubit gates with a depth of $2(n-1)$. Assuming linear qubit connectivity, we use $n-2$ SWAP layers to optimally implement a fully-connected QUBO. A SWAP layer contains \mbox{$\lfloor n/2 \rfloor$} or \mbox{$\lfloor (n-1)/2 \rfloor$} SWAP gates, depending on its parity. This leads to an additional \mbox{$ \Delta N_{\text{line}} = 3 \times \Big\lceil \frac{1}{2} (n-2) \cdot \Big(\lfloor n/2 \rfloor + \lfloor (n-1)/2 \rfloor\Big) \Big\rceil$} entangling two-qubit gates since a SWAP is equivalent to three CNOT gates. When transpiling the final circuit, the overhead gate count becomes $\Delta N_{\text{line}}/3$ due to the cancellation of neighboring entangling gates, i.e. we have only one entangling gate, instead of three, in the addition to the two corresponding to $R_{ZZ}$ in Fig.~\ref{fig:zz_and_swap}. 
The asymptotic factor between circuit gate count on linear and all-to-all connectivity is then
\begin{equation}\label{eq:alpha_QUBO}
    \alpha = 1 + \lim_{n \to \infty} \frac{\Delta N_{\text{line}} /3 }{n(n-1)} = 1.5,
\end{equation}
where we add the overhead of entangling gates on top of the constant factor $1$.
The asymptotic factor between the circuit depth on linear and all-to-all connectivity is then 
\begin{equation}
    \alpha = \lim_{n \to \infty} \frac{3(n-2)}{2(n-1)} = 1.5.
\end{equation}

\figureSeven

\section{Circuit transpilation} \label{sec:transpilation} 

Here, we describe how we transpile QAOA circuits for the higher-order problems. We rely on the `transpile' function of Qiskit~\cite{javadi2024quantum}, which decomposes physical operators into the basis gate set $\{R_{Z}, \sqrt{X}, X, CZ\}$. We assume either the default all-to-all connectivity, or a linear one, by providing a coupling map consisting of a list of connected qubit pairs $[i, i+1]$ and $[i+1, i]$, where $i \in [0, n-1]$. We set the optimization level to 3. Qiskit will thus route the QAOA circuit to the hardware topology with the LightSABER algorithm~\cite{zou2024lightsabre} which is an enhancement of the SABER SWAP algorithm~\cite{li2019mapping}. Next, the instructions in the routed circuit will be decomposed to hardware native ones and undergo further optimizations to remove, e.g., canceling gates, and SWAPs as well as diagonal gates that occur before measurement instructions. 

We now describe the circuit transpilation for the experiments ran on \emph{ibm\_fez} for QAOA for RR3 graphs. First, the qubit connectivity and native gate set are provided by the given backend. We then compile Ansatze circuits with SWAP Strategies as described in the publicly available `qopt-best-practices' repository~\cite{SWAPStrategies_code}. For QUBOs that are not fully connected, the circuit depth can often be reduced by appropriately mapping the decision variables onto the physical qubits. We use the approach of Matsuo et al.~\cite{matsuo2023sat}, which finds an initial mapping by solving multiple satisfiability (SAT) problems to minimize the number of SWAP layers. This method can become time-consuming since SAT problems are NP-hard. For our experiments, we set a timeout time of $20$ seconds. After the circuit is built from $R_{ZZ}$ and SWAP gates, we run the Qiskit transpiler on a low optimization level, i.e. 1, to decompose the gates in $CZ$ gates and simplify any redundant gates.

\section{Weights of quadratized hypergraphs}\label{sec:analytical_weights}

In this section, we derive an analytical formula for the weights $w_{ij}$ of the clique expansion. From Eq.~(\ref{eq:clique_proj}), we can compactly express $w_{ij}$.
Expanding the binomials and eliminating the constant factors results in 
\begin{equation}
     \arg\min_{w_{ij}}  \sum_{ e \in \mathcal{E}: (i,j) \in E(e)} \alpha_e w_{ij}^2 - 2 w_{ij} \alpha_e w_e , \label{eq:analytical_weights_general_2}
\end{equation}
which is now easy to minimize yielding
\begin{equation}
    w_{ij} = \sum_{ e \in \mathcal{E}: (i,j) \in E(e)} w^2_e \Big/ \sum_{ e \in \mathcal{E}: (i,j) \in E(e)} w_e. \label{eq:analytical_weights_general}\\
\end{equation}

Therefore, the weight of the resulting edges is a weighted mean of the hyperedges it belongs to.

The weights of the quadratization resulting from LABS can be more precisely calculated using the fact that all quartic and quadratic terms have weight 2, and 1, respectively. We introduce the variable $I_{ij}$ which is $1$ if $Z_iZ_j$ is in $H_\text{LABS}$ and $0$ otherwise. We also denote by $N_{ij}$ the number of all hyperedges containing edge $(i,j)$. We then obtain from Eq.~(\ref{eq:clique_proj})
\begin{align}
   w_{ij} &=  \arg\min_{w_{ij}}  & \Big( I_{ij} (w_{ij} - 1)^2  \notag \\
   & &+ \sum_{ e \in \mathcal{E}: (i,j) \in E(e)} 2 (w_{ij}-2)^2 \Big) \label{eq:analytical_weights_2} \\
   &= \arg\min_{w_{ij}}   & \Big( w_{ij}^2 ( 2N_{ij} + I_{ij}) \notag \\
   && - 2 w_{ij} ( 4N_{ij} + I_{ij}) \Big) \label{eq:analytical_weights_3}
\end{align}

The edge weights for the quadratized LABS are thus
\begin{equation}
     w_{ij} = 2 - \frac{I_{ij}}{2N_{ij} + I_{ij}} .\label{eq:analytical_weights_4}
\end{equation}
The resulting weights $w_{ij}$ are close to 2 since \mbox{$I_{ij} \in \{ 0, 1\}$} and $N_{ij} \sim O(n)$. To see why $N_{ij} \sim O(n)$, consider another pair of nodes $(h,l)$ with which the edge $(i,j)$ could form a hyperedge $e = (i,j,h,l)$. The expected value of $N_{ij}$ is the sum of the probabilities that the tuple $(i,j,h,l)$ forms a hyperedge in the original hypergraph $\mathcal{E}$, for all pairs of nodes $h,l \in V$, i.e.
\begin{equation}
     \mathbb{E}(N_{ij}) = \sum_{h,l \in V} {\rm Prob}\left[(i,j,h,l)\in\mathcal{E}\right] 
\end{equation}
Since LABS has $O(n^3)$ four-local terms and there are $O(n^4)$ total possible four-local terms, then \mbox{${\rm Prob}\left[(i,j,h,l)\in\mathcal{E}\right]  \sim O(1/n)$} for an arbitrary pair of nodes $(h,l)$. Further, there are $O(n^2)$ ways of choosing the nodes $(h,l)$, thus yielding $\mathbb{E}(N_{ij}) \sim O(n)$.

\section{MPS simulation of random three-regular graphs}\label{sec:RR3_parameters}

The RR3 graph instances we investigate have 40 nodes, which prevents an exact, statevector-based simulation of the underlying quantum circuits without high-performance computing.
Hence, we use a QAOA tailored MPS-based simulator which approximates the energy.
The MPS simulator relies on the methods described in Ref.~\cite{Stoudenmire2020_LimitsQuantum}.
Specifically, the initial state is represented as an MPS with bond dimension 1.
Single-qubit gates are applied by contracting them with the corresponding tensor in the MPS.
Nearest-neighbor two-qubit gates are applied by contracting together the MPS tensors associated with the qubits onto which the gate is applied, contracting the resulting tensor with the gate, and restoring the MPS format through singular value decomposition.
Non-nearest-neighbor two-qubit gates are simulated as they are executed in the hardware, i.e. by using SWAP layers to bring qubits close to each other.
The simulator encodes the MPS in the inverse Vidal form~\cite{Dolgov2020_ParallelTDVP} to avoid tracking the MPS canonization during the simulation.
The quality of the approximation is controlled by the maximum dimension of the tensors composing the MPS, i.e. the bond dimension.
For all our simulations, we set the bond dimension to 20~\cite{gray2018quimb}.

\bibliography{refs}

\end{document}

%% file: allcommands.tex
\usepackage[dvipsnames]{xcolor,colortbl}
\usepackage{xspace,tabularx}
\usepackage{mathtools,amsthm,amssymb, amsmath} 

\usepackage{enumitem} 
\setenumerate[0]{label={\normalfont (\roman*)}}
\usepackage{graphicx}

\usepackage[T1]{fontenc}
\usepackage[UKenglish]{babel}
\usepackage[scaled=0.86]{helvet}
\DeclareMathAlphabet{\mathcal}{OMS}{ntxm}{m}{n}
\usepackage{dsfont} 
\let\mathbb\relax
\let\mathbb\mathds
\usepackage{stmaryrd} 
\makeatletter
\renewcommand{\paragraph}{%
  \@startsection{paragraph}{4}%
  {\z@}{2.25ex \@plus 1ex \@minus .2ex}{-1em}%
  {\normalfont\normalsize\bfseries}%
}
\makeatother
\usepackage{dcolumn}
\usepackage{bm}

\definecolor{ibmblue}{HTML}{0043ce}
\usepackage[colorlinks=true,urlcolor=ibmblue,citecolor=ibmblue,linkcolor=ibmblue]{hyperref}
\usepackage{url}
\usepackage[nameinlink,capitalize,noabbrev]{cleveref}
\crefname{enumi}{Step}{Steps}

\usepackage{physics}
\usepackage{braket}
\usepackage{qcircuit}

%% file: figures.tex
\newcommand{\figureOne}{
    \begin{figure}[htbp]
    \centering
    \includegraphics[width=\columnwidth]{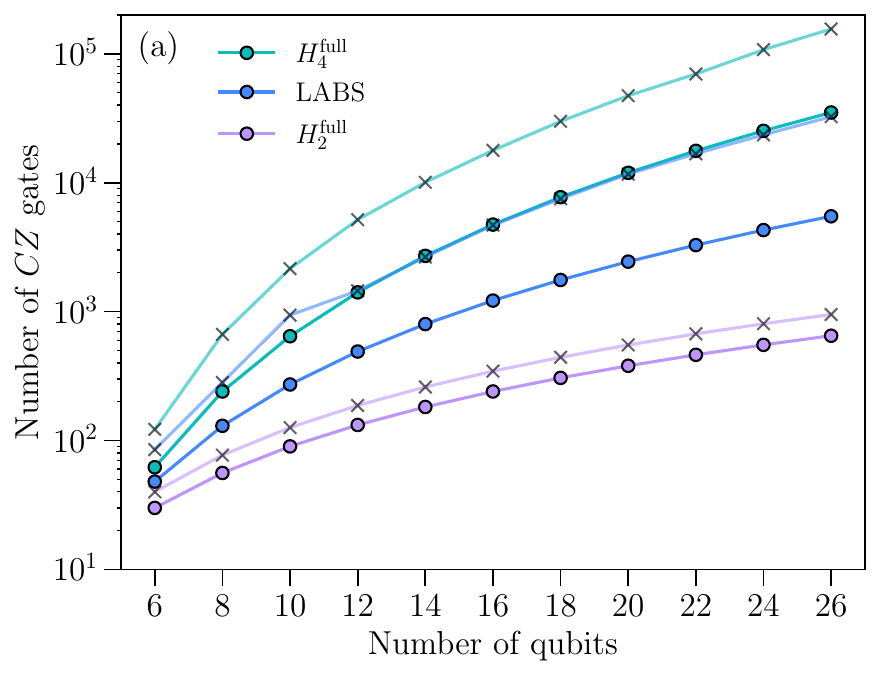} 
    \includegraphics[width=\columnwidth]{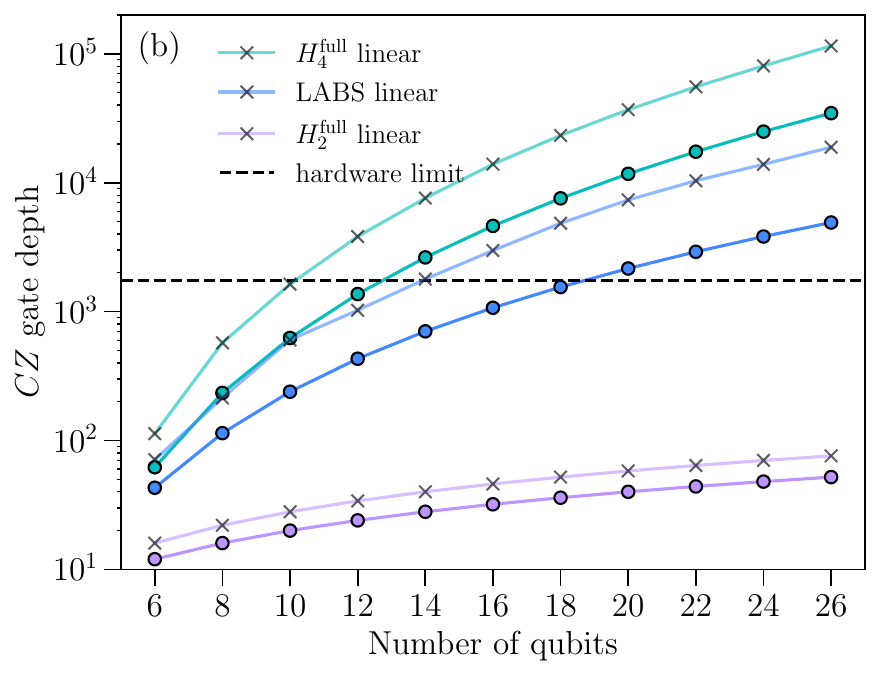}

    \caption[singlelinecheck=false]{Circuit complexity of $e^{-i\gamma H_C}$ for $H_4^{\text{full}}$ (teal) and LABS (blue) compared to a fully-connected QUBO (purple). Panel (a) shows the number of $CZ$ gates as a function of problem size. 
    Panel (b) shows the $CZ$ gate depth of the transpiled circuits. 
    The continuous lines connect raw data points.
    The horizontal dashed line corresponds to the number of $CZ$ gates that fit within the median qubit $T_1$ time on \emph{ibm\_fez}.
    }
    \label{fig:hardness}
\end{figure}
}

\newcommand{\figureTwo}{
    \begin{figure}[htbp]
    \centering
    \includegraphics[width=\linewidth]{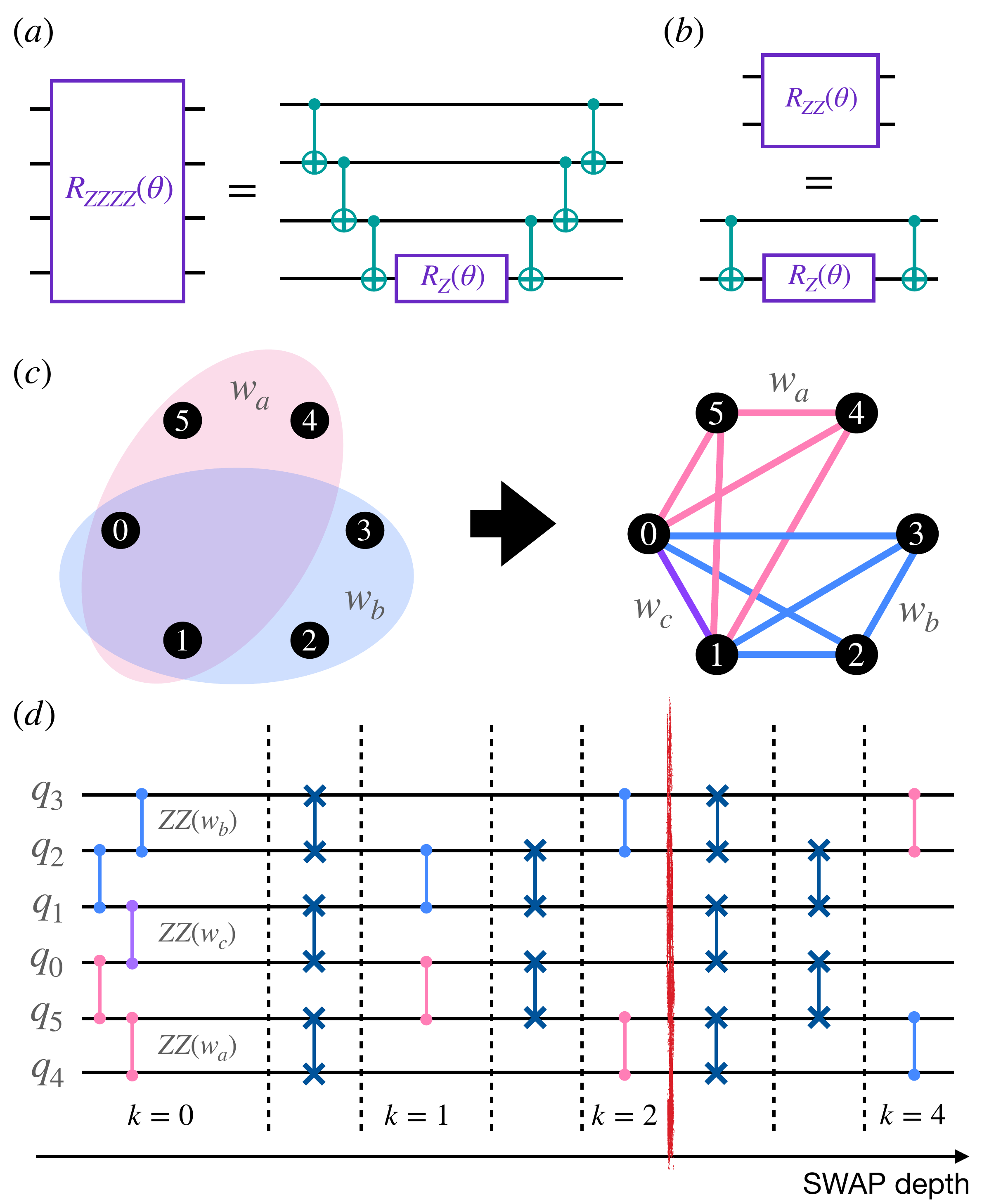} 
    \caption{Construction of shallow Ansatze. (a) and (b) show how a four-local gate $\exp(-i\gamma ZZZZ)$ and a two-local gate $\exp(-i\gamma ZZ)$, respectively, decompose into $R_Z$ and $CX$ gates. (c) The hypergraph clique expansion generates a 2D graph that is easier to implement on quantum hardware. The graph is generated such that each pair of nodes in the original hypergraph has a corresponding edge. Here, the edge $(0,1)$ has weight $w_c = (w_a + w_b) / 2$, whereas the other edges preserve the weights of the single hyperedge they correspondingly belong to.  (d) Circuit implementation of the resulting graph with layers of SWAP gates. To limit the effect of hardware noise, we skip implementing the last two SWAP layers, at the cost losing some correlation with the original problem.
    }
    \label{fig:main_ideas}
\end{figure}
}

\newcommand{\figureThree}{
    \begin{figure}
        \centering
        \includegraphics[width=\columnwidth]{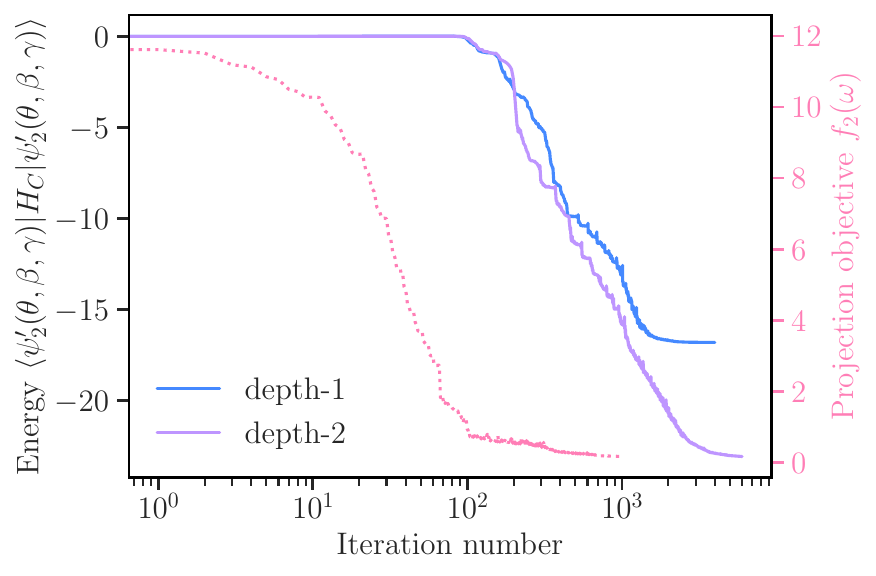} 
        \caption{
        Convergence of the quadratizations as function of COBYLA iterations.
        The clique expansion, i.e., the minimization of Eq.~(\ref{eq:clique_proj}), is shown as the dotted pink curve on the right $y$-axis.
        The solid blue and violet curves show the minimization of Eq.~(\ref{eqn:qaoa_proj}) for QAOA depths one and two, respectively.
        }
        \label{fig:projection_convergence}
    \end{figure}
}

\newcommand{\figureFour}{
    \begin{figure}
        \centering
        \includegraphics[width=\columnwidth]{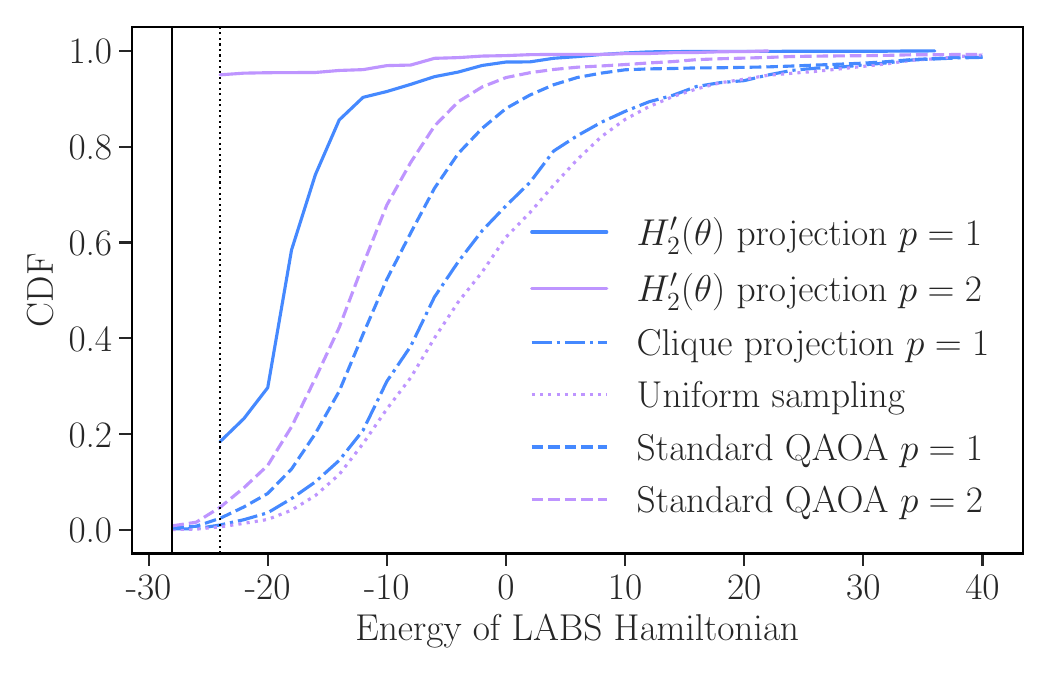}
        \caption{
        Cumulative distribution function of the LABS Hamiltonian energy over $2^{14}$ samples obtained from different quantum circuits.
        The vertical solid line shows the minimum energy found by direct diagonalization.
        The dotted vertical line shows the best solution sampled from the $H_2'(\boldsymbol{\theta})$ quadratization.
        The full energy range of the 12 qubits LABS problem is $[-28, 110]$.
        }
        \label{fig:projection_cdf}
    \end{figure}
}

\newcommand{\figureFive}{
    \begin{figure*}[htbp]
    \centering
    \includegraphics[width=\linewidth]{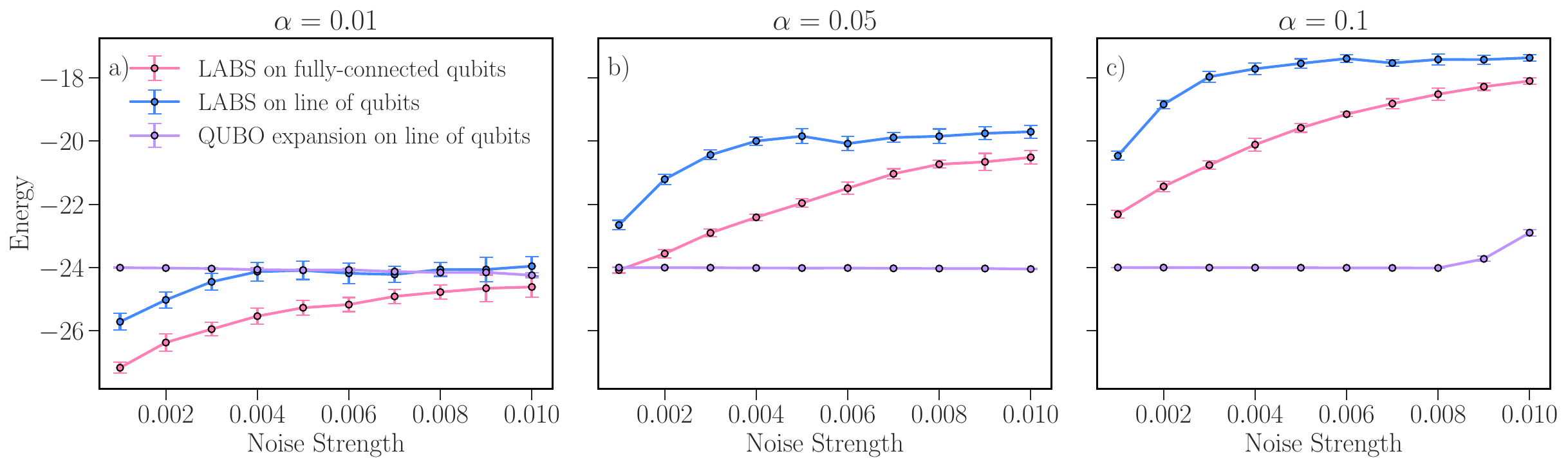} 

    \caption{Evolution of expected performance of depth-two QAOA for LABS versus strength of depolarizing noise for a system of 12 qubits, using the standard Ansatz transpiled on hardware with full connectivity (pink) and with line connectivity (blue) compared to using the quadratized Ansatz transpiled on a line of qubits (purple). The circuit corresponding to the pink line has 600 two-qubit gates, the blue 2094, and the purple 374.
    Each data point is the average of 10 circuit runs and the error bars represent shot noise over 10\,000 samples. 
    }
    \label{fig:simulations}
\end{figure*}
}

\newcommand{\figureSix}{
    \begin{figure}[htbp]
    \centering
    \includegraphics[width=\columnwidth]{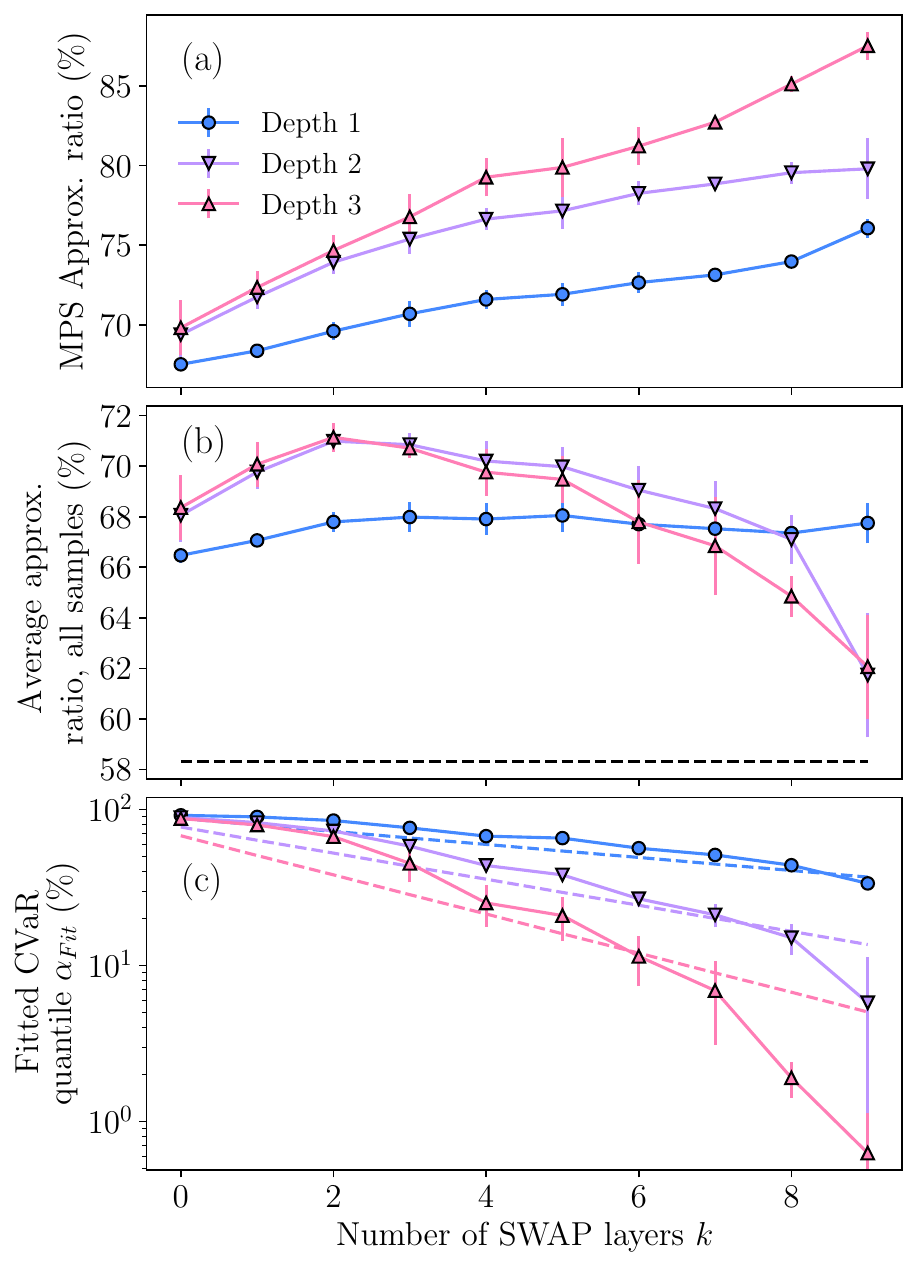}
    \caption{Performance of QAOA as a function of depth and the number of SWAP layers. Each data point is the average of three RR3 graphs that require nine SWAP layers. The error bars show the standard deviation. 
    (a) Approximation ratio of QAOA computed with a MPS evaluation of the energy.
    (b) Measured approximation ratios from 50\,000 samples from \emph{ibm\_fez}.
    The horizontal dashed line is the $58.3\%$ average approximation ratio reported in Ref.~\cite{sack2024blackline}.
    (c) Value of the CVaR quantile $\alpha$ required to make the approximation ratio of the \emph{ibm\_fez} samples in (b) match those of the MPS energy evaluation in (a).
    The dashed lines show a simple theoretical expectation $\alpha_\text{th}$ based on the two-qubit gate fidelity.
    }
    \label{fig:rr3}
\end{figure}
}

\newcommand{\figureSeven}{
    \begin{figure}[htbp]
        \centering
        \includegraphics[width=\linewidth]{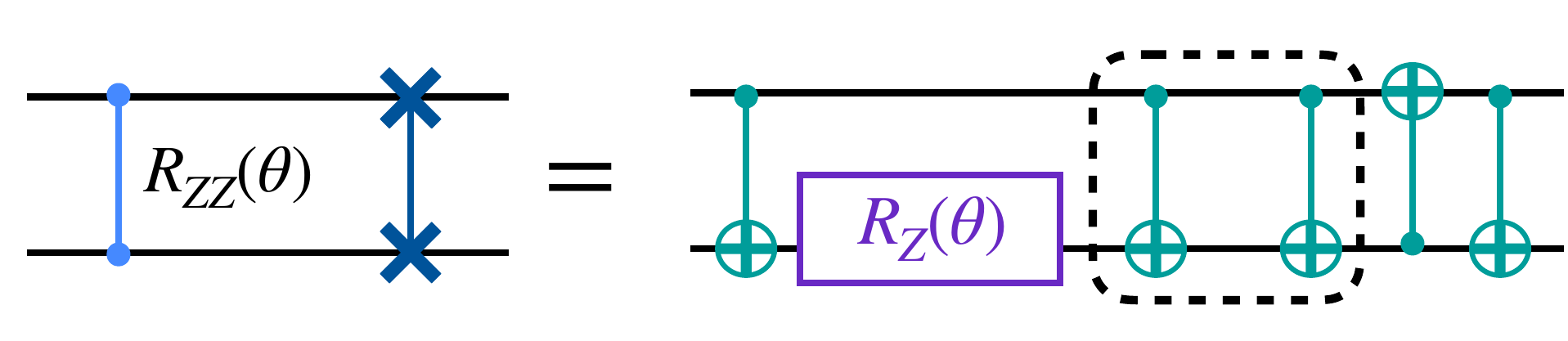} 
        \caption{Decomposition of an $R_{ZZ}$ gate followed by a SWAP gate into a basis gate set made of single-qubit rotations and a $CX$. Here, two consecutive $CX$ gates cancel. This construction is typical in SWAP networks on linear connectivity, see Fig.~\ref{fig:main_ideas}(d).}
        \label{fig:zz_and_swap}
    \end{figure}
}

%% file: refs.bib
@conference{romero2024bias,
  title={{Bias-field digitized counterdiabatic quantum algorithm for higher-order binary optimization}},
  author={Romero, Sebasti{\'a}n V and Visuri, Anne-Maria and Cadavid, Alejandro Gomez and Solano, Enrique and Hegade, Narendra N},
  year={2024},
eprint = "2409.04477",
    archivePrefix = "arXiv",
}

@inproceedings{li2019mapping, author = {Li, Gushu and Ding, Yufei and Xie, Yuan}, title = {{Tackling the Qubit Mapping Problem for NISQ-Era Quantum Devices}}, year = {2019}, isbn = {9781450362405}, publisher = {Association for Computing Machinery}, address = {New York, NY, USA}, url = {https://doi.org/10.1145/3297858.3304023}, doi = {10.1145/3297858.3304023}, booktitle = {Proceedings of the Twenty-Fourth International Conference on Architectural Support for Programming Languages and Operating Systems}, pages = {1001–1014}, numpages = {14}, location = {Providence, RI, USA}, series = {ASPLOS '19} 
}

@conference{zou2024lightsabre,
  title={{LightSABRE: A Lightweight and Enhanced SABRE Algorithm}},
  author={Zou, Henry and Treinish, Matthew and Hartman, Kevin and Ivrii, Alexander and Lishman, Jake},
  year={2024},
    eprint = "2409.08368",
    archivePrefix = "arXiv",
}

@conference{koch2025decathlon,
    author = "Koch, Thorsten and Bernal Neira, David E. and Chen, Ying and Cortiana, Giorgio and Egger, Daniel J. and Heese, Raoul and Hegade, Narendra N. and Cadavid, Alejandro Gomez and Huang, Rhea and Itoko, Toshinari and others",
    title = "{{Quantum Optimization Benchmark Library -- The Intractable Decathlon}}",
    eprint = "2504.03832",
    archivePrefix = "arXiv",
    month = "4",
    year = "2025"
}

@misc{herrman2021multiangle,
      title={Multi-angle Quantum Approximate Optimization Algorithm}, 
      author={Rebekah Herrman and Phillip C. Lotshaw and James Ostrowski and Travis S. Humble and George Siopsis},
      year={2021},
      eprint={2109.11455},
      archivePrefix={arXiv},
      primaryClass={quant-ph},
      url={https://arxiv.org/abs/2109.11455}, 
}

@inproceedings{shor1994algorithms,
  title={Algorithms for quantum computation: discrete logarithms and factoring},
  author={Shor, Peter W},
  booktitle={Proceedings 35th annual symposium on foundations of computer science},
  pages={124–134},
  year={1994},
  organization={IEEE},
  doi={10.1109/SFCS.1994.365700}
}

@article{peruzzo2014variational,
  title={A variational eigenvalue solver on a photonic quantum processor},
  author={Peruzzo, Alberto and McClean, Jarrod and Shadbolt, Peter and Yung, Man-Hong and Zhou, Xiao-Qi and Love, Peter J and Aspuru-Guzik, Al{\'a}n and O’Brien, Jeremy L},
  journal={Nat. Commun.},
  volume={5},
  number={1},
  pages={4213},
  year={2014},
  publisher={Nature Publishing Group UK London},
  doi={https://doi.org/10.1038/ncomms5213}
}

@article{kandala2017hardware,
  title={Hardware-efficient variational quantum eigensolver for small molecules and quantum magnets},
  author={Kandala, Abhinav and Mezzacapo, Antonio and Temme, Kristan and Takita, Maika and Brink, Markus and Chow, Jerry M and Gambetta, Jay M},
  journal={Nature},
  volume={549},
  number={7671},
  pages={242–246},
  year={2017},
  publisher={Nature Publishing Group},
  doi ={https://doi.org/10.1038/nature23879}
}

@article{tilly2022variational,
  title={The {Variational} {Quantum} {Eigensolver}: a review of methods and best practices},
  author={Tilly, Jules and Chen, Hongxiang and Cao, Shuxiang and Picozzi, Dario and Setia, Kanav and Li, Ying and Grant, Edward and Wossnig, Leonard and Rungger, Ivan and Booth, George H and others},
  journal={Phys. Rep.},
  volume={986},
  pages={1–128},
  year={2022},
  publisher={Elsevier},
  doi = {https://doi.org/10.1016/j.physrep.2022.08.003}
}

@article{cao2018potential,
  author={Cao, Y. and Romero, J. and Aspuru-Guzik, A.},
  journal={IBM J. Res. Dev.}, 
  title={Potential of quantum computing for drug discovery}, 
  year={2018},
  volume={62},
  number={6},
  pages={6:1–6:20},
  doi={10.1147/jrd.2018.2888987}
}

@article{blunt2022perspective,
  title={Perspective on the {Current} {State}-of-the-{Art} of {Quantum} {Computing} for {Drug} {Discovery} {Applications}},
  author={Blunt, Nick S and Camps, Joan and Crawford, Ophelia and Izs{\'a}k, R{\'o}bert and Leontica, Sebastian and Mirani, Arjun and Moylett, Alexandra E and Scivier, Sam A and Sunderhauf, Christoph and Schopf, Patrick and others},
  journal={J. Chem. Theory Comput.},
  volume={18},
  number={12},
  pages={7001–7023},
  year={2022},
  publisher={ACS Publications},
doi = {https://doi.org/10.1021/acs.jctc.2c00574}
}

@article{holmes2022connecting,
  title={Connecting {Ansatz} {Expressibility} to {Gradient} {Magnitudes} and {Barren} {Plateaus}},
  author={Holmes, Zo{\"e} and Sharma, Kunal and Cerezo, Marco and Coles, Patrick J},
  journal={Phys. Rev. X Quantum},
  volume={3},
  number={1},
  pages={010313},
  year={2022},
  publisher={APS},
doi = {https://doi.org/10.1103/Phys. Rev. XQuantum.3.010313},
}

@book{peterson2007computer,
	title = {Computer {Networks}},
    author = {Peterson, L. Larry and Davie, S. Bruce},
    publisher = {Elsevier},
	isbn = {978-0-12-818200-0},
	url = {https://shop.elsevier.com/books/computer-networks/peterson/978-0-12-818200-0},
	urldate = {2025-03-28},
	month = oct,
	year = {2020},
}

@book{markowitz2008portfolio,
 ISBN = {9780300013726},
 URL = {http://www.jstor.org/stable/j.ctt1bh4c8h},
 author = {Harry M. Markowitz},
 publisher = {Yale University Press},
 title = {Portfolio Selection: Efficient Diversification of Investments},
 urldate = {2025-03-28},
 year = {1959}
}

@article{egger2020quantum,
  title={Quantum {Computing} for {Finance}: {State}-of-the-{Art} and {Future} {Prospects}},
  author={Egger, Daniel J and Gambella, Claudio and Marecek, Jakub and McFaddin, Scott and Mevissen, Martin and Raymond, Rudy and Simonetto, Andrea and Woerner, Stefan and Yndurain, Elena},
  journal={IEEE Trans. Quantum Eng.},
  volume={1},
  pages={1–24},
  year={2020},
  publisher={IEEE},
  doi={10.1109/TQE.2020.3030314},
}

@article{domino2022quadratic,
  title={Quadratic and higher-order unconstrained binary optimization of railway rescheduling for quantum computing},
  author={Domino, Krzysztof and Kundu, Akash and Salehi, {\"O}zlem and Krawiec, Krzysztof},
  journal={Quantum Inf. Process.},
  volume={21},
  number={9},
  pages={337},
  year={2022},
  publisher={Springer},
doi = {https://doi.org/10.1007/s11128-022-03670-y}
}

@article{pascariu2024formulation,
  title={Formulation of train routing selection problem for different real-time traffic management objectives},
  author={Pascariu, Bianca and Sama, Marcella and Pellegrini, Paola and D’Ariano, Andrea and Rodriguez, Joaquin and Pacciarelli, Dario},
  journal={J. Rail Transp. Plan. Manag.},
  volume={31},
  pages={100460},
  year={2024},
  publisher={Elsevier},
doi = {https://doi.org/10.1016/j.jrtpm.2024.100460}
}

@article{sbihi2010combinatorial,
  title={Combinatorial optimization and {Green} {Logistics}},
  author={Sbihi, Abdelkader and Eglese, Richard W},
  journal={Ann. Oper. Res.},
  volume={175},
  pages={159–175},
  year={2010},
  publisher={Springer},
    doi = {https://doi.org/10.1007/s10479-009-0651-z}
}

@article{karp1975computational,
  title={On the {Computational} {Complexity} of {Combinatorial} {Problems}},
  author={Karp, Richard M},
  journal={Networks},
  volume={5},
  number={1},
  pages={45–68},
  year={1975},
  publisher={Wiley Online Library},
doi = {https://doi.org/10.1002/net.1975.5.1.45}
}

@article{zuckerman1996unapproximable,
  title={On {Unapproximable} {Versions} of {{NP}-Complete} {Problems}},
  author={Zuckerman, David},
  journal={SIAM J. Comput.},
  volume={25},
  number={6},
  pages={1293–1304},
  year={1996},
  publisher={SIAM},
doi = {https://doi.org/10.1137/S0097539794266407}
}

@article{abbas2024challenges,
  title={Challenges and opportunities in quantum optimization},
  author={Abbas, Amira and Ambainis, Andris and Augustino, Brandon and B{\"a}rtschi, Andreas and Buhrman, Harry and Coffrin, Carleton and Cortiana, Giorgio and Dunjko, Vedran and Egger, Daniel J and Elmegreen, Bruce G and others},
  journal={Nat. Rev. Phys.},
  pages={718–735},
volume = {6},
  year={2024},
  publisher={Nature Publishing Group UK London},
doi = {https://doi.org/10.1038/s42254-024-00770-9}
}

@conference{farhi2014quantum,
    title={A {Quantum} {Approximate} {Optimization} {Algorithm}},
    author={Farhi, Edward and Goldstone, Jeffrey and Gutmann, Sam},
year = {2014},
    eprint={1411.4028},
    archivePrefix={arXiv},
}

@article{blekos2024review,
  title={A review on {Quantum} {Approximate} {Optimization} {Algorithm} and its variants},
  author={Blekos, Kostas and Brand, Dean and Ceschini, Andrea and Chou, Chiao-Hui and Li, Rui-Hao and Pandya, Komal and Summer, Alessandro},
  journal={Phys. Rep.},
  volume={1068},
  pages={1–66},
  year={2024},
  publisher={Elsevier},
doi = {https://doi.org/10.1016/j.physrep.2024.03.002}
}

@article{finnila1994quantum,
  title={Quantum annealing: {A} new method for minimizing multidimensional functions},
  author={Finnila, Aleta Berk and Gomez, Maria A and Sebenik, C and Stenson, Catherine and Doll, Jimmie D},
  journal={Chem. Phys. Lett.},
  volume={219},
  number={5-6},
  pages={343–348},
  year={1994},
  publisher={Elsevier},
doi = {https://doi.org/10.1016/0009-2614(94)00117-0}
}

@article{rajak2023quantum,
  title={Quantum annealing: an overview},
  author={Rajak, Atanu and Suzuki, Sei and Dutta, Amit and Chakrabarti, Bikas K},
  journal={Philos. Trans. R. Soc., A},
  volume={381},
  number={2241},
  pages={20210417},
  year={2023},
  publisher={The Royal Society},
doi = {https://doi.org/10.1098/rsta.2021.0417}
}

@article{zhou2020quantum, 
  title={{Quantum} {Approximate} {Optimization} {Algorithm}: Performance, {Mechanism}, and {Implementation} on {Near-Term} {Devices}},
  author={Zhou, Leo and Wang, Sheng-Tao and Choi, Soonwon and Pichler, Hannes and Lukin, Mikhail D},
  journal={Phys. Rev. X},
  volume={10},
  number={2},
  pages={021067},
  year={2020},
  publisher={APS},
    doi = {https://doi.org/10.1103/PhysRevX.10.021067}
}

@article{brandhofer2022benchmarking,
  title={Benchmarking the performance of portfolio optimization with {QAOA}},
  author={Brandhofer, Sebastian and Braun, Daniel and Dehn, Vanessa and Hellstern, Gerhard and H{\"u}ls, Matthias and Ji, Yanjun and Polian, Ilia and Bhatia, Amandeep Singh and Wellens, Thomas},
  journal={Quantum Inf. Process.},
  volume={22},
  number={1},
  pages={25},
  year={2022},
  publisher={Springer},
    doi = {https://doi.org/10.1007/s11128-022-03766-5}
}

@article{willsch2020benchmarking,
  title={Benchmarking the quantum approximate optimization algorithm},
  author={Willsch, Madita and Willsch, Dennis and Jin, Fengping and De Raedt, Hans and Michielsen, Kristel},
  journal={Quantum Inf. Process.},
  volume={19},
  pages={1–24},
  year={2020},
  publisher={Springer},
doi = {https://doi.org/10.1007/s11128-020-02692-8}
}

@article{tate2023warm,
  title={{Warm-{Started}} {QAOA} with {Custom} {Mixers} {Provably} {Converges} and {Computationally} {Beats} {{Goemans}-{Williamson}}'s {Max-{Cut}} at {Low} {Circuit} {Depths}},
  author={Tate, Reuben and Moondra, Jai and Gard, Bryan and Mohler, Greg and Gupta, Swati},
  journal={Quantum},
  volume={7},
  pages={1121},
  year={2023},
  publisher={Verein zur F{\"o}rderung des Open Access Publizierens in den Quantenwissenschaften},
doi = {https://doi.org/10.22331/q-2023-09-26-1121}
}

@article{egger2021warm,
  title={Warm-starting quantum optimization},
  author={Egger, Daniel J and Mare{\v{c}}ek, Jakub and Woerner, Stefan},
  journal={Quantum},
  volume={5},
  pages={479},
  year={2021},
  publisher={Verein zur F{\"o}rderung des Open Access Publizierens in den Quantenwissenschaften},
doi = {https://doi.org/10.22331/q-2021-06-17-479}
}

@article{rehfeldt2023faster,
  title={Faster exact solution of sparse {MaxCut} and {QUBO} problems},
  author={Rehfeldt, Daniel and Koch, Thorsten and Shinano, Yuji},
  journal={Math. Program. Comput.},
  volume={15},
  number={3},
  pages={445–470},
  year={2023},
  publisher={Springer},
doi = {https://doi.org/10.1007/s12532-023-00236-6}
}

@article{goemans1995improved,
  title={Improved approximation algorithms for maximum cut and satisfiability problems using semidefinite programming},
  author={Goemans, Michel X and Williamson, David P},
  journal={J. ACM},
  volume={42},
  number={6},
  pages={1115–1145},
  year={1995},
  publisher={ACM New York, NY, USA},
doi = {https://doi.org/10.1145/227683.227684}
}

@article{gamarnik2021overlap,
  title={The overlap gap property: A topological barrier to optimizing over random structures},
  author={Gamarnik, David},
  journal={Proc. Natl. Acad. Sci. U. S. A.},
  volume={118},
  number={41},
  pages={e2108492118},
  year={2021},
  publisher={National Acad Sciences},
doi = {https://doi.org/10.1073/pnas.2108492118}
}

@inproceedings{lykov2023fast,
  title={Fast {Simulation} of {{High}-{Depth}} {QAOA} {Circuits}},
  author={Lykov, Danylo and Shaydulin, Ruslan and Sun, Yue and Alexeev, Yuri and Pistoia, Marco},
  booktitle={Proceedings of the SC'23 Workshops of The International Conference on High Performance Computing, Network, Storage, and Analysis},
  pages={1443–1451},
  year={2023},
doi = { https://doi.org/10.1145/3624062.3624216}
}

@article{lykov2023sampling,
  title={Sampling frequency thresholds for the quantum advantage of the quantum approximate optimization algorithm},
  author={Lykov, Danylo and Wurtz, Jonathan and Poole, Cody and Saffman, Mark and Noel, Tom and Alexeev, Yuri},
  journal={Npj Quantum Inf.},
  volume={9},
  number={1},
  pages={73},
  year={2023},
  publisher={Nature Publishing Group UK London},
doi = {https://doi.org/10.1038/s41534-023-00718-4}
}

@article{akshay2020reachability,
  title={Reachability {Deficits} in {Quantum} {Approximate} {Optimization}},
  author={Akshay, Vishwanathan and Philathong, Hariphan and Morales, Mauro ES and Biamonte, Jacob D},
  journal={Phys. Rev. Lett.},
  volume={124},
  number={9},
  pages={090504},
  year={2020},
  publisher={APS},
doi = {https://doi.org/10.1103/PhysRevLett.124.090504}
}

@article{brest2021low,
  title={Low {Autocorrelation} {Binary} {Sequences}: {{Best}-{Known}} {Peak} {Sidelob}e {Level} {Values}},
  author={Brest, Janez and Bo{\v{s}}kovi{\'c}, Borko},
  journal={IEEE Access},
  volume={9},
  pages={67713–67723},
  year={2021},
  publisher={IEEE},
doi={10.1109/ACCESS.2021.3077541}
}

@article{packebusch2016low,
  title={Low autocorrelation binary sequences},
  author={Packebusch, Tom and Mertens, Stephan},
  journal={J. Phys. A: Math. Theor.},
  volume={49},
  number={16},
  pages={165001},
  year={2016},
  publisher={IOP Publishing},
doi = {10.1088/1751-8113/49/16/165001}
}

@article{shaydulin2024evidence,
  title={Evidence of scaling advantage for the quantum approximate optimization algorithm on a classically intractable problem},
  author={Shaydulin, Ruslan and Li, Changhao and Chakrabarti, Shouvanik and DeCross, Matthew and Herman, Dylan and Kumar, Niraj and Larson, Jeffrey and Lykov, Danylo and Minssen, Pierre and Sun, Yue and others},
  journal={Sci. Adv.},
  volume={10},
  number={22},
  pages={eadm6761},
  year={2024},
  publisher={American Association for the Advancement of Science},
doi = {10.1126/sciadv.adm6761}
}

@misc{gurobi,
  title = {Gurobi {Optimization} {LLC.}},
  howpublished = {\url{https://www.gurobi.com}},
}

@misc{cplex,
  title = {{IBM} {ILOG} {CPLEX. V20.1}: User’s manual for cplex},
  journal = {International Business Machines Corporation.},
    howpublished = {\url{https://www.ibm.com/docs/en/icos/22.1.2?topic=optimizers-users-manual-cplex}}
}

@article{lucas2014ising,
  title={Ising formulations of many {NP} problems},
  author={Lucas, Andrew},
  journal={Front. Phys.},
  volume={2},
  pages={5},
  year={2014},
  publisher={Frontiers Media SA},
doi = { https://doi.org/10.3389/fphy.2014.00005}
}

@article{maciejewski2024design,
  title={Design and execution of quantum circuits using tens of superconducting qubits and thousands of gates for dense {Ising} optimization problems},
  author={Maciejewski, Filip B and Hadfield, Stuart and Hall, Benjamin and Hodson, Mark and Dupont, Maxime and Evert, Bram and Sud, James and Alam, M Sohaib and Wang, Zhihui and Jeffrey, Stephen and others},
  journal={Phys. Rev. A},
  volume={22},
  number={4},
  pages={044074},
  year={2024},
  publisher={APS},
doi  = {https://doi.org/10.1103/PhysRevApplied.22.044074}
}

@article{zhu2022adaptive,
  title={Adaptive quantum approximate optimization algorithm for solving combinatorial problems on a quantum computer},
  author={Zhu, Linghua and Tang, Ho Lun and Barron, George S and Calderon-Vargas, F A and Mayhall, Nicholas J and Barnes, Edwin and Economou, Sophia E},
  journal={Phys. Rev. Res.},
  volume={4},
  number={3},
  pages={033029},
  year={2022},
  publisher={APS},
doi = {https://doi.org/10.1103/PhysRevResearch.4.033029}
}

@article{fuchs2022constraint,
  title={Constraint {Preserving} {Mixers} for the {Quantum} {Approximate} {Optimization} {Algorithm}},
  author={Fuchs, Franz Georg and Lye, Kjetil Olsen and Nilsen, Halvor M{\o}ll and Stasik, Alexander Johannes and Sartor, Giorgio},
  journal={Algorithms},
  volume={15},
  number={6},
  pages={202},
  year={2022},
  publisher={Multidisciplinary Digital Publishing Institute},
doi = {https://doi.org/10.3390/a15060202}
}

@article{he2023alignment,
  title={Alignment between initial state and mixer improves {QAOA} performance for constrained optimization},
  author={He, Zichang and Shaydulin, Ruslan and Chakrabarti, Shouvanik and Herman, Dylan and Li, Changhao and Sun, Yue and Pistoia, Marco},
  journal={Npj Quantum Inf.},
  volume={9},
  number={1},
  pages={121},
  year={2023},
  publisher={Nature Publishing Group UK London},
doi = {https://doi.org/10.1038/s41534-023-00787-5}
}

@conference{niu2019optimizing,
  title={Optimizing {QAOA}: Success {Probability} and {Runtime} {Dependence} on {Circuit} {Depth}},
  author={Niu, Murphy Yuezhen and Lu, Sirui and Chuang, Isaac L},
year = {2019},
    eprint={1905.12134},
    archivePrefix={arXiv},
}

@article{golay977LABS,
  author={Golay, M.},
  journal={IEEE Trans. Inf. Theory}, 
  title={Sieves for low autocorrelation binary sequences}, 
  year={1977},
  volume={23},
  number={1},
  pages={43-51},
  doi={10.1109/TIT.1977.1055653}
}

@article{bernasconi1987title,
  author    = {J. Bernasconi},
  title     = {Low autocorrelation binary sequences : statistical mechanics and configuration space analysis},
  year      = {1987},
  volume    = {48},
  number    = {4},
  pages     = {559--567},
journal   = {J. Phys.},
  url       = {https://jphys.journaldephysique.org/articles/jphys/abs/1987/04/jphys_1987__48_4_559_0/jphys_1987__48_4_559_0.html},
}

@conference{javadi2024quantum,
  title={Quantum computing with {Qiskit}},
  author={Javadi-Abhari, Ali and Treinish, Matthew and Krsulich, Kevin and Wood, Christopher J and Lishman, Jake and Gacon, Julien and Martiel, Simon and Nation, Paul D and Bishop, Lev S and Cross, Andrew W and others},
year = {2024},
    eprint={2405.08810},
    archivePrefix={arXiv},
}

@article{rosenhaus2019introduction,
  title={An introduction to the {SYK} model},
  author={Rosenhaus, Vladimir},
  journal={J. Phys. A: Math. Theor.},
  volume={52},
  number={32},
  pages={323001},
  year={2019},
  publisher={IOP Publishing},
doi = {10.1088/1751-8121/ab2ce1}
}

@article{luo2019quantum,
  title={Quantum simulation of the non-fermi-liquid state of {{Sachdev}-{Ye}-{Kitaev}} model},
  author={Luo, Zhihuang and You, Yi-Zhuang and Li, Jun and Jian, Chao-Ming and Lu, Dawei and Xu, Cenke and Zeng, Bei and Laflamme, Raymond},
  journal={Npj Quantum Inf.},
  volume={5},
  number={1},
  pages={53},
  year={2019},
  publisher={Nature Publishing Group UK London},
doi = {https://doi.org/10.1038/s41534-019-0166-7}
}

@article{kobrin2021many,
  title={{Many-{body}} {Chaos} in the {{Sachdev}-{Ye}-{Kitaev}} {Model}},
  author={Kobrin, Bryce and Yang, Zhenbin and Kahanamoku-Meyer, Gregory D and Olund, Christopher T and Moore, Joel E and Stanford, Douglas and Yao, Norman Y},
  journal={Phys. Rev. Lett.},
  volume={126},
  number={3},
  pages={030602},
  year={2021},
  publisher={APS},
doi = {https://doi.org/10.1103/PhysRevLett.126.030602}
}

@article{matsuo2023sat,
  title={A {SAT} {Approach} to the {Initial} {Mapping} {Problem} in {SWAP} {Gate} {Insertion} for {Commuting} {Gates}},
  author={Matsuo, Atsushi and Yamashita, Shigeru and Egger, Daniel J},
  journal={IEICE Trans. Fundam. Electron. Commun. Comput. Sci.},
  volume={106},
  number={11},
  pages={1424--1431},
  year={2023},
  publisher={The Institute of Electronics, Information and Communication Engineers},
doi = {https://doi.org/10.1587/transfun.2022EAP1159}
}

@inproceedings{mandal2020compressed,
  title={Compressed quadratization of higher order binary optimization problems},
  author={Mandal, Avradip and Roy, Arnab and Upadhyay, Sarvagya and Ushijima-Mwesigwa, Hayato},
  booktitle={Proceedings of the 17th ACM International Conference on Computing Frontiers},
  pages={126--131},
  year={2020},
doi = {https://doi.org/10.1145/3387902.3392627}
}

@inproceedings{suppakitpaisarn2024utilizing,
  title={Utilizing {Graph} {Sparsification} for {{Pre}-processing} in {Max} {Cut} {QUBO} {Solver}},
  author={Suppakitpaisarn, Vorapong and Hao, Jin-Kao},
  booktitle={Metaheuristics International Conference},
  pages={219--233},
  year={2024},
  organization={Springer},
    doi = {https://doi.org/10.1007/978-3-031-62912-9_22}
}

@article{sack2024blackline,
  title = {{Large-scale} quantum approximate optimization on nonplanar graphs with machine learning noise mitigation},
  author = {Sack, Stefan H. and Egger, Daniel J.},
  journal = {Phys. Rev. Res.},
  volume = {6},
  issue = {1},
  pages = {013223},
  numpages = {13},
  year = {2024},
  month = {Mar},
  publisher = {American Physical Society},
  doi = {10.1103/PhysRevResearch.6.013223},
  url = {https://link.aps.org/doi/10.1103/PhysRevResearch.6.013223}
}

@article{weidenfeller2202scaling,
  title={Scaling of the quantum approximate optimization algorithm on superconducting qubit based hardware},
  author={Weidenfeller, Johannes and Valor, Lucia C and Gacon, Julien and Tornow, Caroline and Bello, Luciano and Woerner, Stefan and Egger, Daniel J},
  year={2022},
  journal={Quantum},
    volume = {6},
    pages = {870},
    doi = {https://doi.org/10.22331/q-2022-12-07-870}
}

@article{Harrigan_2021,
   title={Quantum approximate optimization of non-planar graph problems on a planar superconducting processor},
   volume={17},
   ISSN={1745-2481},
   url={http://dx.doi.org/10.1038/s41567-020-01105-y},
   DOI={10.1038/s41567-020-01105-y},
   number={3},
   journal={Nat. Phys.},
   publisher={Springer Science and Business Media LLC},
   author={Harrigan, Matthew P. and Sung, Kevin J. and Neeley, Matthew and Satzinger, Kevin J. and Arute, Frank and Arya, Kunal and Atalaya, Juan and Bardin, Joseph C. and Barends, Rami and Boixo, Sergio and others},
   year={2021},
   month=feb, pages={332–336} 
}

@article{barkoutsos2020cvar,
   title={Improving {Variational} {Quantum} {Optimization} using {CVaR}},
   volume={4},
   ISSN={2521-327X},
   url={http://dx.doi.org/10.22331/q-2020-04-20-256},
   DOI={10.22331/q-2020-04-20-256},
   journal={Quantum},
   publisher={Verein zur Forderung des Open Access Publizierens in den Quantenwissenschaften},
   author={Barkoutsos, Panagiotis Kl. and Nannicini, Giacomo and Robert, Anton and Tavernelli, Ivano and Woerner, Stefan},
   year={2020},
   month=apr, pages={256} 
}

@article{barron2024cvar,
   title={Provable bounds for {noise-free} expectation values computed from noisy samples},
   volume={4},
   ISSN={2662-8457},
   url={http://dx.doi.org/10.1038/s43588-024-00709-1},
   DOI={10.1038/s43588-024-00709-1},
   number={11},
   journal={Nat. Comput. Sci.},
   publisher={Springer Science and Business Media LLC},
   author={Barron, Samantha V. and Egger, Daniel J. and Pelofske, Elijah and Bärtschi, Andreas and Eidenbenz, Stephan and Lehmkuehler, Matthis and Woerner, Stefan},
   year={2024},
   month=nov, pages = "865–875" }

@conference{mckay2023benchmarking,
      title={{Benchmarking Quantum Processor Performance at Scale}}, 
      author={David C. McKay and Ian Hincks and Emily J. Pritchett and Malcolm Carroll and Luke C. G. Govia and Seth T. Merkel},
year = {2023},
    eprint={2311.05933},
    archivePrefix={arXiv},
}

@conference{larocca2024review,
  title={A {Review} of {Barren} {Plateaus} in {Variational} {Quantum} {Computing}},
  author={Larocca, Martin and Thanasilp, Supanut and Wang, Samson and Sharma, Kunal and Biamonte, Jacob and Coles, Patrick J and Cincio, Lukasz and McClean, Jarrod R and Holmes, Zo{\"e} and Cerezo, M},
note = {},
year = {2024},
    eprint={2405.00781},
    archivePrefix={arXiv},
}

@article{arrasmith2022equivalence,
  title={Equivalence of quantum barren plateaus to cost concentration and narrow gorges},
  author={Arrasmith, Andrew and Holmes, Zo{\"e} and Cerezo, Marco and Coles, Patrick J},
  journal={Quantum Sci. Technol.},
  volume={7},
  number={4},
  pages={045015},
  year={2022},
  publisher={IOP Publishing},
doi = {10.1088/2058-9565/ac7d06}
}

@article{mcclean2018barren,
  title={Barren plateaus in quantum neural network training landscapes},
  author={McClean, Jarrod R and Boixo, Sergio and Smelyanskiy, Vadim N and Babbush, Ryan and Neven, Hartmut},
  journal={Nat. Commun.},
  volume={9},
  number={1},
  pages={4812},
  year={2018},
  publisher={Nature Publishing Group UK London},
doi = {https://doi.org/10.1038/s41467-018-07090-4}
}

@article{bittel2021training,
  title={Training {Variational} {Quantum} {Algorithms} is {NP}-hard},
  author={Bittel, Lennart and Kliesch, Martin},
  journal={Phys. Rev. Lett.},
  volume={127},
  number={12},
  pages={120502},
  year={2021},
  publisher={APS},
doi = {https://doi.org/10.1103/PhysRevLett.127.120502}
}

@misc{SWAPStrategies_code,
  title = {Quantum optimization best practices},
  publisher = {GitHub},
  journal = {GitHub repository},
  howpublished = {\url{https://github.com/qiskit-community/qopt-best-practices}},
}

@article{sack2023recursive,
  title={Recursive greedy initialization of the quantum approximate optimization algorithm with guaranteed improvement},
  author={Sack, Stefan H and Medina, Raimel A and Kueng, Richard and Serbyn, Maksym},
  journal={Phys. Rev. A},
  volume={107},
  number={6},
  pages={062404},
  year={2023},
  publisher={APS},
doi = { https://doi.org/10.1103/PhysRevA.107.062404}
}

@article{gray2018quimb,
  title={quimb: A python package for quantum information and many-body calculations},
  author={Gray, Johnnie},
  journal={J. Open Source Softw.},
  volume={3},
  number={29},
  pages={819},
  year={2018},
doi = {https://joss.theoj.org/papers/10.21105/joss.00819}
}

@article{barkoutsos2018quantum,
  title={Quantum algorithms for electronic structure calculations: Particle-hole Hamiltonian and optimized wave-function expansions},
  author={Barkoutsos, Panagiotis Kl and Gonthier, Jerome F and Sokolov, Igor and Moll, Nikolaj and Salis, Gian and Fuhrer, Andreas and Ganzhorn, Marc and Egger, Daniel J and Troyer, Matthias and Mezzacapo, Antonio and others},
  journal={Phys. Rev. A},
  volume={98},
  number={2},
  pages={022322},
  year={2018},
  publisher={APS},
doi = {https://doi.org/10.1103/PhysRevA.98.022322}
}

@article{santra2024squeezing,
  title={Squeezing and quantum approximate optimization},
  author={Santra, Gopal Chandra and Jendrzejewski, Fred and Hauke, Philipp and Egger, Daniel J},
  journal={Phys. Rev. A},
  volume={109},
  number={1},
  pages={012413},
  year={2024},
  publisher={APS},
doi = {https://doi.org/10.1103/PhysRevA.109.012413}
}

@article{brest2018heuristic,
  title={A heuristic algorithm for a low autocorrelation binary sequence problem with odd length and high merit factor},
  author={Brest, Janez and Bo{\v{s}}kovi{\'c}, Borko},
  journal={IEEE Access},
  volume={6},
  pages={4127–4134},
  year={2018},
  publisher={IEEE},
doi = {10.1109/ACCESS.2018.2789916}
}

@article{Stoudenmire2020_LimitsQuantum,
  title = {What Limits the Simulation of Quantum Computers?},
  author = {Zhou, Yiqing and Stoudenmire, E. Miles and Waintal, Xavier},
  journal = {Phys. Rev. X},
  volume = {10},
  issue = {4},
  pages = {041038},
  numpages = {15},
  year = {2020},
  month = {Nov},
  publisher = {American Physical Society},
  doi = {10.1103/PhysRevX.10.041038},
  url = {https://link.aps.org/doi/10.1103/PhysRevX.10.041038}
}

@article{Dolgov2020_ParallelTDVP,
  title = {Parallel time-dependent variational principle algorithm for matrix product states},
  author = {Secular, Paul and Gourianov, Nikita and Lubasch, Michael and Dolgov, Sergey and Clark, Stephen R. and Jaksch, Dieter},
  journal = {Phys. Rev. B},
  volume = {101},
  issue = {23},
  pages = {235123},
  numpages = {17},
  year = {2020},
  month = {Jun},
  publisher = {American Physical Society},
  doi = {10.1103/PhysRevB.101.235123},
  url = {https://link.aps.org/doi/10.1103/PhysRevB.101.235123}
}
